\documentclass[fleqn,usenatbib]{rasti}
\usepackage{xcolor}
\usepackage{comment}
\newcommand{\hii}{{\sc H\,ii}}
\usepackage{fontawesome}
\newcommand{\nii}{[{\sc N\,ii}]}
\newcommand{\sii}{[{\sc S\,ii}]}
\newcommand{\NGC}{NGC\,}

\newcommand\textref[1]{\textbf{\color{blue} #1}}
\definecolor{viridian}{rgb}{0.165, 0.498, 0.384}

\usepackage{newtxtext,newtxmath}
\usepackage[T1]{fontenc}
\usepackage{textgreek}
\DeclareRobustCommand{\VAN}[3]{#2}
\let\VANthebibliography\thebibliography
\def\thebibliography{\DeclareRobustCommand{\VAN}[3]{##3}\VANthebibliography}

\usepackage{graphicx}	% Including figure files
\usepackage{amsmath}	% Advanced maths commands
\usepackage{caption}
\usepackage{subcaption}
\title{Reconstructing Robust Background IFU spectra using Machine Learning}

\author[Carter L. Rhea]{
Carter Lee Rhea,$^{1,2}$\thanks{E-mail: carter.rhea@umontreal.ca}
Julie Hlavacek-Larrondo,$^{1,3}$
Justine Giroux,$^{4}$ 
Auriane Thilloy,$^{1,3}$
Hyunseop Choi,$^{1,3,6}$\newauthor
Laurie Rousseau-Nepton,$^{7,8,9}$
Marie-Lou Gendron-Marsolais,$^{10}$
Mario Pasquato,$^{1,3,5,6}$
Simon Prunet$^{11}$
\\
$^{1}$Département de Physique, Université de Montréal, Succ. Centre-Ville, Montréal, Québec, H3C 3J7, Canada\\
$^{2}$Centre de Recherche en Astrophysique du Québec (CRAQ), Québec, QC, G1V 0A6, Canada\\
$^{3}$ Ciela, Computation and Astrophysical Data Analysis Institute, Montreal, Quebec, Canada\\
$^{4}$ Département de Vision Numérique, Université Laval, Québec, QC, H3A 1B9, Canada\\
$^{5}$ Physics and Astronomy Department Galileo Galilei, University of Padova, Vicolo dell’Osservatorio 3, I–35122, Padova\\
$^{6}$ Mila - Quebec Artificial Intelligence Institute, Montreal, Quebec, Canada\\
$^{7}$ Canada-France-Hawaii Telescope, 65-1238 Mamalahoa Hwy, Kamuela, Hawaii 96743, USA\\
$^{8}$ David A. Dunlap Department of Astronomy \& Astrophysics, University of Toronto, 50 St. George Street, Toronto, ON M5S 3H4, Canada\\
$^{9}$  Dunlap Institute for Astronomy \& Astrophysics, University of Toronto, 50 St. George Street, Toronto, ON M5S 3H4, Canada\\
$^{10}$ Instituto de Astrofísica de Andalucía, IAA-CSIC, Apartado 3004, 18080 Granada, España\\
$^{11}$Université Côte d'Azur, Observatoire de la Côte d'Azur, CNRS, Laboratoire Lagrange, France\\
}

\date{Accepted XXX. Received YYY; in original form ZZZ}

\pubyear{2023}

\begin{document}
\label{firstpage}
\pagerange{\pageref{firstpage}--\pageref{lastpage}}
 \maketitle

\begin{abstract}

%Accurately selecting a background model is critical for spectroscopic analysis. Incorrect background spectra result in inaccurate flux and kinematic measurements, particularly in regions not dominated by source emissions. 
In astronomy, spectroscopy consists of observing an astrophysical source and extracting its spectrum of electromagnetic radiation. Once extracted, a model is fit to the spectra to measure the observables, leading to an understanding of the underlying physics of the emission mechanism. One crucial, and often overlooked, aspect of this model is the background emission, which contains foreground and background astrophysical sources, intervening atmospheric emission, and artifacts related to the instrument such as noise.
This paper proposes an algorithmic approach to constructing a background model for SITELLE observations using statistical tools and supervised machine learning algorithms. 
SITELLE is an imaging Fourier Transform Spectrometer located at the Canada-France-Hawai'i Telescope, which produces a 3-dimensional data cube containing the position of the emission (2 dimensions) and the spectrum of the emission. SITELLE has a wide field of view (11 arcminutes by 11 arcminutes), which makes the background emission particularly challenging to model.
We apply a segmentation algorithm implemented in \texttt{photutils} to divide the data cube into background and source spaxels. After applying a principal component analysis (PCA) on the background spaxels, we train an artificial neural network to interpolate from the background to the source spaxels in the PCA coefficient space, which allows us to generate a local background model over the entire data cube.
We highlight the performance of this methodology by applying it to SITELLE observations obtained of a SIGNALS galaxy, \NGC4449, and the Perseus galaxy cluster of galaxies, NGC 1275. We discuss the physical interpretation of the principal components and noise reduction in the resulting PCA-based reconstructions. Additionally, we compare the fit results using our new background modeling approach to standard methods used in the literature and find that our method better captures the emission from \hii{} regions in NGC 4449 and the faint emission regions in NGC 1275. These methods also demonstrate that the background does change as a function of the position of the datacube.
While the approach is applied explicitly to SITELLE data in this study, we argue that it can be readily adapted to any integral field unit (IFU) style data, enabling the user to obtain more robust measurements on the flux of the emission lines. Finally, we note that the SITELLE data analysis pipeline \href{https://github.com/crhea93/LUCI}{\faicon{github}\texttt{LUCI}} now contains an optimized implementation of the methodology developed in this paper.
\end{abstract}

% Include between one and six keywords.

%%%%%%%%%%%%%%%%%%%%%%%%%%%%%%%%%%%%%%%%%%%%%%%%%%

%%%%%%%%%%%%%%%%% BODY OF PAPER %%%%%%%%%%%%%%%%%%

\section{Introduction}

Integral Field Units (IFUs) are rapidly changing our understanding of galaxies (e.g., \citealt{law_observing_2015}; \citealt{emsellem_phangs-muse_2022}; \citealt{barbosa_geminigmos_2009}; \citealt{edwards_diverse_2009}; \citealt{groves_phangs-muse_2023}). Due to their complementary spatial and spectral coverage, IFUs enable the study of spectra as a function of complex morphologies. However, extracting key parameters, such as the flux from strong emission lines, absorption lines, and continuum emission, requires careful consideration of the background emission. This problem is exacerbated in many IFU studies of nearby galaxies and galaxy clusters since the astrophysical objects of interest can take up the majority of the instrument's field of view (FOV), thus rendering the background difficult to determine. Moreover, the background emission in IFUs with a large FOV, such as MUSE WFM (\citealt{bacon_muse_2017}) and SITELLE (\citealt{drissen_sitelle_2019}), is not homogeneous over the data cube. 

The background is often decomposed into four primary components: the skylines and any background/foreground emission source in addition to detector noise. Additionally, a galaxy's stellar continuum can contribute negatively to the overall flux measurements.
Stellar continuum is regularly modeled using a combination of GANDALF/PPXF modeling (\citealt{sarzi_gandalf_2017}; \citealt{cappellari_ppxf_2012}). 
This requires multiple (and strong) absorption components that are frequently unavailable (such as in SITELLE observations that typically only cover a small bandpass). Therefore, we need a robust method to model the background.

A standard technique employed in the past consists of taking a large background region without any point sources nor emission from the astrophysical source and calculating the mean spectrum of this region; a large background region is taken to reduce the overall noise level. This assumes homogeneity of the background over the entire FOV of the instrument. 
While this works well for capturing skylines that are stable over the FOV, it neglects additional components such as intervening sources and/or the stellar continuum. Moreover, the skylines can vary strongly over larger FOVs depending on the observing conditions, so this method could fail. 
While numerous background detection algorithms exist, such as matched-filter analysis, they often depend on \textit{a priori} knowledge of background region spectra (\citealt{masias_review_2012}; \citealt{ramella_finding_2001}; \citealt{bacon_muse_2017}; \citealt{hopkins_new_2002}). Matched-filter analysis takes a library of sample spectra and tries to find which template best fits the data by minimizing the Euclidean distance between them. Matched-filter analysis and similar methodologies require a broad library of template spectra covering any physical or astrophysical phenomena that may appear in the background. Thus, this method is impractical to implement. 

The SITELLE instrument on the Canada-France-Hawai'i Telescope boasts a field of view of 11 arcminutes $\times$ 11 arcminutes (e.g., \citealt{drissen_sitelle_2019}). SITELLE is an imaging Fourier transform spectrograph, meaning that in addition to taking a two-dimensional image of a target, it also captures a spectrum over a bandpass for each pixel, generating a three-dimensional datacube where the third axis is the spectrum.
SITELLE contains over 4 million spaxels representing source and/or background emission. 
SITELLE's unique design allows the observer to change the resolving power, R$=\frac{\lambda}{\Delta\lambda}$, where $\lambda$ is the wavelength and $\Delta\lambda$ is the wavelength step. The spectral resolution of SITELLE ranges from R$\sim1$ up to R$\sim10,000$.
The SIGNALS (Star-formation, Ionized Gas and Nebular Abundances Legacy Survey; \citealt{rousseau-nepton_signals_2019}) project uses SITELLE to target over 50,000 spatially-resolved \hii{}-regions in more than 40 galaxies (D$\leq40$ Mpc). One of the most challenging steps in tackling these observations has been the accurate estimation of the background emission, as most galaxies observed in SIGNALS cover the entire field of view.

This paper presents a novel approach to modeling the background using a combination of classic image segmentation algorithms, principal component analysis (PCA), and artificial neural networks. In $\S$\ref{sec:methods}, we outline each component of the algorithm. 
In $\S$\ref{sec:results}, we discuss how the algorithm  can be applied to SITELLE observations of two different environments: the dwarf galaxy NGC 4449 and the massive elliptical galaxy 1275.
%we discuss how the algorithm can be applied to a galaxy, NGC 4449, and a galaxy cluster, NGC 1275. 
These two observations were chosen because they contrast each other and demonstrate this algorithm's use in studying SIGNALS' galaxies and other SITELLE targets. Both targets are detailed in sections \ref{sec:NGC4449} and \ref{sec:NGC1275}.  In $\S$\ref{sec:discussion}, we compare this methodology to standard background modeling techniques, discuss potential modifications to the algorithm depending on use cases, and explore its potential in other fields of astronomy and applications to other IFU-like instruments.

\section{Observations and Methods}\label{sec:methods}
%In this section, we start by outlining  the steps in the proposed methodology, followed by a detailed discussion of each step.
 In this section, we start by describing the SITELLE observations chosen for the testing of our algorithm. We next explain the community's traditional method of background modeling. We then outline the steps to obtain a background model in the proposed methodology, followed by a detailed discussion of each step.

\subsection{Observations}
\subsubsection{NGC 4449}
\NGC4449 was observed by SITELLE using the SN3 (6510-6850 \AA) filter centered on RA 12h28m09.8s and DEC +44°05m51.2s (P.I. Rousseau-Nepton; R$\approx5000$) as part of the SIGNALS project. The SN3 filter at this redshift contains strong emission lines present in \hii{} regions, diffuse ionized gas (DIG), planetary nebulae, and supernova remnants: \nii{}$\lambda$6548, \nii{}$\lambda$6583, H\textalpha{}, \sii{}$\lambda$6716, and \sii{}$\lambda$6731. SITELLE samples the object at approximately 5.9 parsecs per pixel (0.32 arcseconds) due to the proximity of the dwarf galaxy (z$=0.00069$). 
The SITELLE observations studied in this paper have been preprocessed using the standard SITELLE preprocessing pipeline known as ORBS (\textit{Outil de	Réduction Binoculaire pour SITELLE}; \citealt{martin_calibrations_2017}). ORBS reduces SITELLE raw observations by calibrating the raw CCD images, aligning the datacubes from the two cameras on SITELLE, combining the interferometric data from the cameras, applying a phase correction to the combined cube, calculating the Fourier Transformation of the combined cube, and applying a final wavelength and flux calibration to the combined cube.
Once the datacube has been reduced, we create a deep image of the observation (shown in figure \ref{fig:deepImages} a). The deep image is created by summing over the spectral axis, creating a 2-dimensional image where each pixel value corresponds to the total flux in a pixel. 
The data analysis (including the background modeling, subtract, and emission-line fitting) is completed with \texttt{LUCI} -- an analysis pipeline developed specifically for SITELLE (\citealt{rhea_luci_2021}). \texttt{LUCI} is a general-purpose fitting pipeline that uses machine learning to reduce computation time while increasing computational accuracy. Using pre-trained neural networks, \texttt{LUCI} initializes fit parameters such as the velocity offset and broadening of emission lines, resulting in accurate solutions and robust flux measurements.

\subsubsection{NGC 1275}
 
In figure \ref{fig:deepImages} b, we display the SN3 deep image of NGC 1275 (z=0.0179) as observed by SITELLE. NGC 1275 was observed in 2016 as part of the science verification stage of SITELLE (P.I. G. Morrison). Since the observation was part of the science verification phase, the resolution of the cube was set to R$\sim$1800. The data were initially reduced using the ORBS software (\citealt{martin_calibrations_2017}) and analyzed using \texttt{LUCI} (\citealt{rhea_luci_2021}). Just as with NGC 4449, the SN3 filter targets the following emission lines: \nii{}$\lambda$6548, \nii{}$\lambda$6583, H\textalpha{}, \sii{}$\lambda$6716, and \sii{}$\lambda$6731.

 \subsection{Traditional Background Modeling}
Previously, two different methodologies existed to model the background emission in a data cube from optical IFUs targeting extra-galactic sources. The first consisted of taking a representative background region\footnote{A representative background region is considered a region in the datacube sufficiently distant from the primary source of emission containing no point sources} and extracting the mean spectrum. In doing so, we capture the average amount of sky emission and minimize the noise. This representative background spectrum is then subtracted from each spaxel before fitting. This technique has been used extensively (e.g.  \citealt{alcorn_extended_2023}; \citealt{gendron-marsolais_revealing_2018}; \citealt{groves_phangs-muse_2023}), but, contrary to the methodology described in this paper, it assumes homogeneity across the FOV, which is not guaranteed since the amplitude of the skylines and/or the properties of the DIG for an extended source can change over the FOV. Similar strategies for choosing the pixels consisting of the background are proposed (see \cite{rousseau-nepton_signals_2019} or Vigneron et al. 2023 for examples) though these also assume relative homogeneity.

The second contemporary methodology extracts the background from individual emission regions. It relies on first pinpointing individual source emission regions, defining an aperture encircling the source emission, masking out the source region, and taking the median spectrum from this region (i.e., \citealt{moumen_3d_2019}; \citealt{schroetter_muse_2016}; \citealt{jones_sdss_2017}; \citealt{law_data_2016}; \citealt{bundy_overview_2015}). Although this method creates very accurate backgrounds and skirts the problem of global homogeneity since it constructs local background models, it requires an algorithm that accurately identifies emission regions. Furthermore, it is most accurate when the emission regions are sufficiently disentangled spatially from one another to ensure an accurate background measurement. In the case of NGC4449, this method is not applicable due to the tangled morphology of individual \hii{} regions. Similarly, the complex morphology of the emission in NGC 1275 makes this methodology impractical.
 
\subsection{An Overview of the Algorithm}
The following steps describe our proposed background spectrum methodology:
\begin{enumerate}
    \item Use a segmentation algorithm to isolate the background pixels from the source pixels.
    \item Apply PCA to construct a subspace representing the background components.
    \item Project each background spaxel into a truncated PCA space.
    \item 2D interpolation on the masked (source) pixels into the truncated PCA space.
\end{enumerate}

Following these steps, we isolate spaxels corresponding to the background, construct a reduced-order representation of the background, and impute the model in masked regions.

\subsection{Segmentation Algorithm}\label{sec:segAlgo}
%\carter{The SITELLE observations studied in this paper have been preprocessed using the standard SITELLE preprocessing pipeline known as ORBS (\textit{Outil de	Réduction Binoculaire pour SITELLE}; \citealt{martin_calibrations_2017}). ORBS reduces SITELLE raw observations by calibrating the raw CCD images, aligning the datacubes from the two cameras on SITELLE, combining the interferometric data from the cameras, applying a phase correction to the combined cube, calculating the Fourier Transformation of the combined cube, and applying a final wavelength and flux calibration to the combined cube.
%Once the datacube has been reduced, we create a deep image of the observation (shown in figure \ref{fig:deepImages}). The deep image is created by summing over the spectral axis, creating a 2-dimensional image where each pixel value corresponds to the total flux in a pixel. 

In order to construct a model of the background spectra, we must determine which pixels are associated with background emission versus source emission, where source emission includes emission from the foreground (or background) point sources, galaxies, and/or the astrophysical phenomena being studied. For example, a SITELLE data cube initially taken to study \hii{} regions in a nearby galaxy will contain contaminating stars, planetary nebulae, and supernova remnants that should not be included in the background model, as well as the hydrogen emission we aim to study. 
%While numerous background detection algorithms exist, such as matched-filter analysis, they often depend on \textit{a-priori} knowledge of source region spectra (\citealt{masias_review_2012}; \citealt{ramella_finding_2001}; \citealt{bacon_muse_2017}; \citealt{hopkins_new_2002}). 
Here, instead of making any assumptions about the source spectra, we use the image segmentation algorithm from \texttt{photutils} on the SITELLE deep image, which is used to find extended and point sources. Therefore, the segmentation algorithm sees only the net flux in each pixel.  We then apply the \href{https://photutils.readthedocs.io/en/stable/segmentation.html}{\texttt{photutils}} image segmentation implementation (\citealt{bradley_astropyphotutils_2023}). 
\
\begin{figure*}
    \centering
    \includegraphics[width=0.95\textwidth]{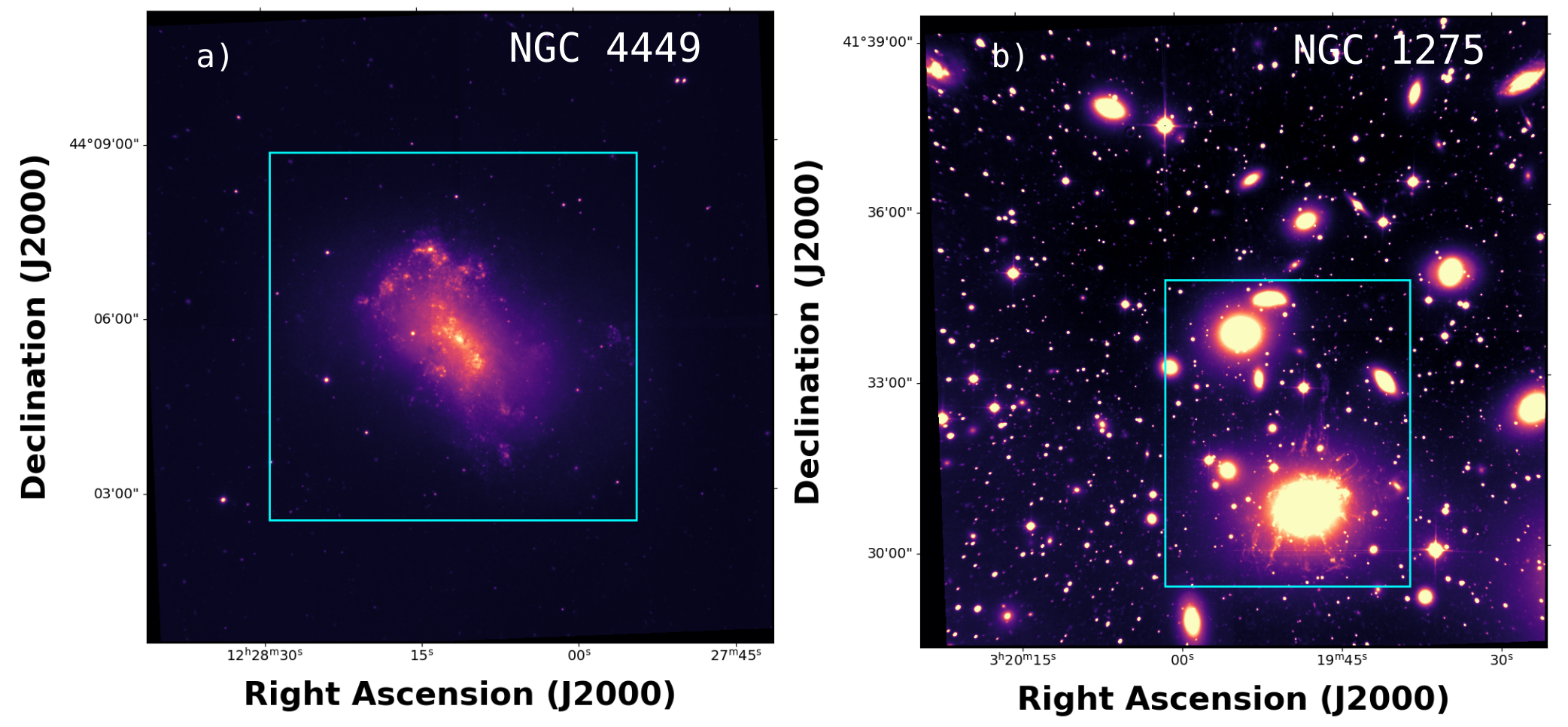}
    \caption{Deep image of the SN3 filter observation of \NGC4449 (a) and NGC 1275 (b) taken with SITELLE. The color scale was chosen to highlight the omnipresent DIG emission. Background and foreground objects appear as bright point sources in the image surrounding the galaxy. The cyan square represents the area over which the segmentation map was applied.}
    \label{fig:deepImages}
\end{figure*}

\texttt{photutils} detects sources in an image by applying an image segmentation algorithm. The first step in the segmentation algorithm is to create a rough estimate of the background. To do so, the image is gridded into subregions of a given box size - this is a user-chosen parameter. The box size is chosen so that it is larger than the scale of sources in the image and small enough to capture background variations. We have experimentally found a box size of 50$\times$50 pixels to work well for two SITELLE datacubes imaging nearby galaxies. Pixels in a single grid box are convolved with a Gaussian kernel; we select a size of 3$\times$3 following the recommendation of \texttt{photutils}. Then, the background level and background rms (root mean square) are calculated for each cell using the sigma-clipped median background. This background map is then interpolated to match the original size of the image and subtracted from the image. The background-subtracted deep image is then convolved with a 3$\times$3 gaussian in order to reduce the noise in the resulting image. Finally, we detect sources that are above a user-defined threshold using \texttt{photutils.detect\_sources}. We set the threshold to be 0.01 times the background noise rms level. This produces a map of source regions, and, conversely, background regions, in the deep image\footnote{These steps follow the standard procedure outlined at \href{https://photutils.readthedocs.io/en/stable/segmentation.html}{https://photutils.readthedocs.io/en/stable/segmentation.html}.}.
%We show the segmentation maps for NGC 4449 and NGC 1275 in figures \ref{fig:backgroundMapsCombined} a and b, respectively.
%\textref{The source pixels are shown in color and the background pixels are shown in black.}
Our methodology suffers from similar issues to the local background method discussed above if the segmentation algorithm does not accurately disentangle the background from the source emission.
%The background and background noise images are created using \texttt{photutils.Background2D} with a box size of 50$\times$50, a 3$\times$3 kernel, and the \texttt{photutils.background.BackgroundMedian} algorithm. These values were selected experimentally. We then subtracted the background and applied a threshold of 0.01 times the background noise root-mean-square. Subsequently, the background-subtracted deep image is convolved with a 3$\times$3 Gaussian kernel. Thereafter, we apply \texttt{photutils.detect\_sources} on the smoothed image produced in the previous step\footnote{These steps follow the standard procedure outlined at \href{https://photutils.readthedocs.io/en/stable/segmentation.html}{https://photutils.readthedocs.io/en/stable/segmentation.html}}. 

\begin{figure*}
    \centering
    \includegraphics[width=0.95\textwidth]{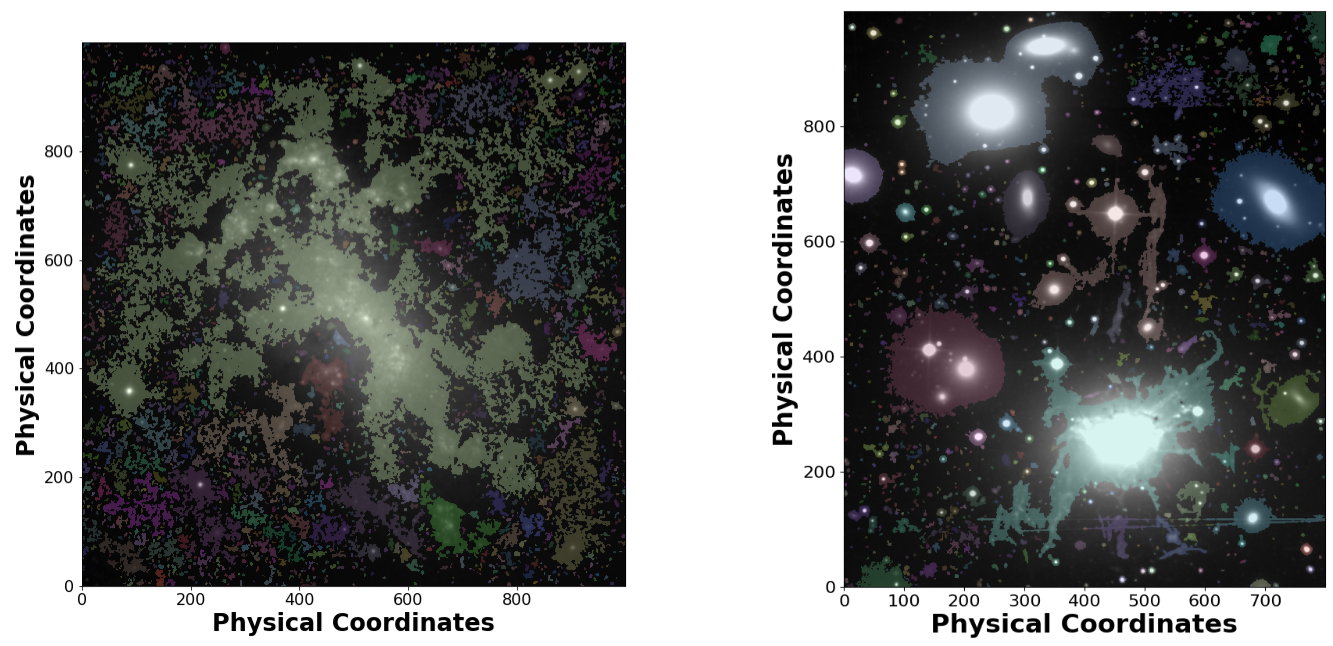}
    \caption{Segmentation map created from the deep image of \NGC4449 (a) and NGC 1275 (b). Each colored region represents a segment found by \texttt{photutils}. %Colors are recycled, so each colored region does not necessarily represent a distinct segment. Note that the segmentation of each colored region has no physical representation and is an artifact of the algorithm.
    The colors in the segmentation maps are randomly selected and assigned to each region. Each pixel of the same color is considered by the segmentation algorithm as belonging to the same region. For our purposes, this has no meaning other than colored pixels are source regions, and background pixels are shown in black.
    }
    \label{fig:backgroundMapsCombined}
\end{figure*}

\subsection{Principal Component Analysis}
Following the creation of the background pixel mask in the previous section, we construct a vector sub-space that represents the ensemble of background spectra in a reduced-dimensional space. In order to accomplish this, we apply a PCA (e.g., \citealt{jolliffe_principal_2005}; \citealt{abdi_principal_2010}; \citealt{bro_principal_2014}; \citealt{ringner_what_2008}). 
We apply a dimensionality reduction technique to the background spectra in order to capture the important \textref{spectral} features and remove as much noise as possible from the background.
PCA decomposes data into vectors that contain the maximum amount of variance describing the data. PCA is typically framed in the following manner:
\begin{enumerate}
    \item Calculate the mean background spectrum. %\mario{sometimes people also calculate the standard deviation and divide by it; whether you do it or not you could say it and say why}
    \item After subtracting the mean from each spectrum, calculate the covariance matrix of the features.
    \item Calculate the singular value decomposition of the data. Then, obtain the eigenvalues of the covariance matrix by taking the square of the singular values.
    \item Construct a transformation matrix with the \texttt{k}-eigenvectors.
\end{enumerate}
Mathematically, we can reconstruct any spectrum, $s_r$, using a linear combination of the principal components, $p$, and the coefficients, $\alpha$, unique to a given spaxel and the mean spectrum $\mu$:
\begin{equation}
    s_{r} = \mu + \sum_{i=0}^k \alpha_i p_i.
\end{equation}
Each principal component covers the entire wavelength coverage of the spaxel.
A user-defined hyperparameter, $\texttt{k}$, indicates the number of principal components to retain; in this manner, PCA acts as a dimensionality-reduction technique. Additionally, since each eigenvalue is sorted from greatest to the least, the first eigenvector (or eigenspectrum) holds the most variance, the second eigenspectrum holds the second-most variance, and so on. Therefore, eigenspectra of higher orders contain little to no variance and can be discarded as noise. PCA has been used extensively in the literature for these reasons to study galactic spectra (i.e., \citealt{ronen_principal_1999}; \citealt{bailey_principal_2012}; \citealt{heyer_application_1997}; \citealt{yip_spectral_2004}; \citealt{smith_probing_2022}; \citealt{mcgurk_principal_2010}). In this work, we apply the incremental PCA implementation from \href{https://scikit-learn.org/stable/modules/generated/sklearn.decomposition.PCA.html}{\texttt{scikit-learn}} (\citealt{pedregosa_scikit-learn_2011}). 

\begin{figure*}
    \centering
    \includegraphics[width=0.95\textwidth]{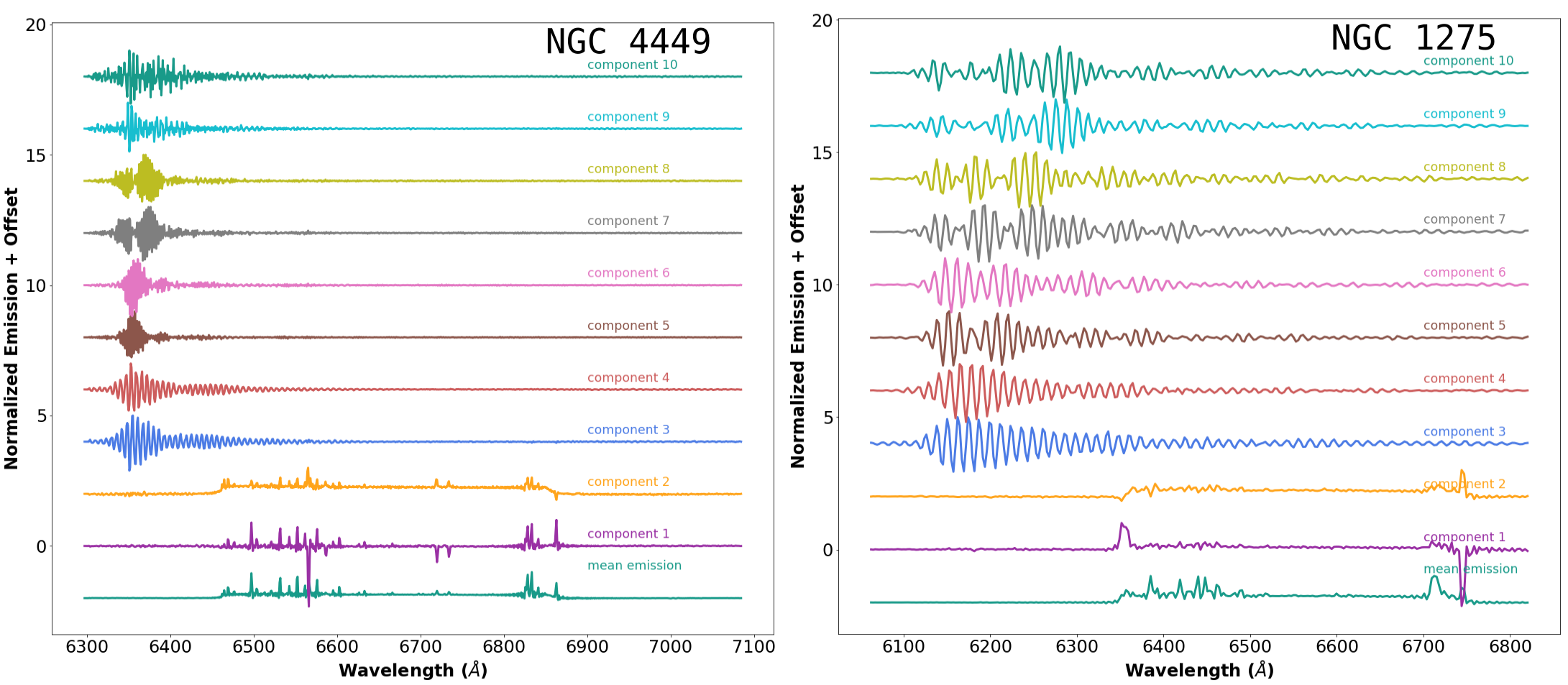}
    \caption{The first ten principal components of the background spaxels in \NGC4449 (left) and NGC 1275 (right) as identified in figure \ref{fig:backgroundMapsCombined}. Each spectrum represents a different component, including the mean background emission. The spectra have been normalized to unity in this figure in order to highlight the major features of each. All higher-order components (meaning components above 10) contain uniquely noise signatures and have been left out of this representation for readability. The majority of the sky-line emission shows up in the mean emission for both observations.}
    \label{fig:PCAcombined}
\end{figure*}

Before applying the PCA, we apply two normalizations since, if we do not apply normalization, the PCA may select a feature to be more important than the others based only on its scale rather than the actual variance it explains. We first normalize each spectrum by the maximum value in the spectrum between 670 and 675 nanometers  so that we can easily scale the source pixels as well. We chose this part of the spectrum since there are no strong emission lines, and it is dominated by noise while also encapsulating the continuum level. Since PCA works best with normalized values, we further normalize each spectrum by its maximum value. We do not restrict the wavelength over which the PCA is conducted.

The result of constructing a principal component-based subspace are \texttt{k}-eigenspectra and their corresponding coefficients for each spaxel in the background space.
%(we show the eigenvectors for NGC 4449 and NGC 1275 in figure \ref{fig:PCAcombined} and the scree plots in figure \ref{fig:screePlots}; the scree plot shows the explained variance (i.e. the amount of variance explained by a single component) as a function of the principal component).
Thus, we can assign each pixel a \texttt{k}-dimensional coefficient vector. To use this subspace to make background spectra for masked source pixels, we can interpolate in our \texttt{k}-dimensional space.

\begin{figure*}
    \centering
    \includegraphics[width=0.98\textwidth]{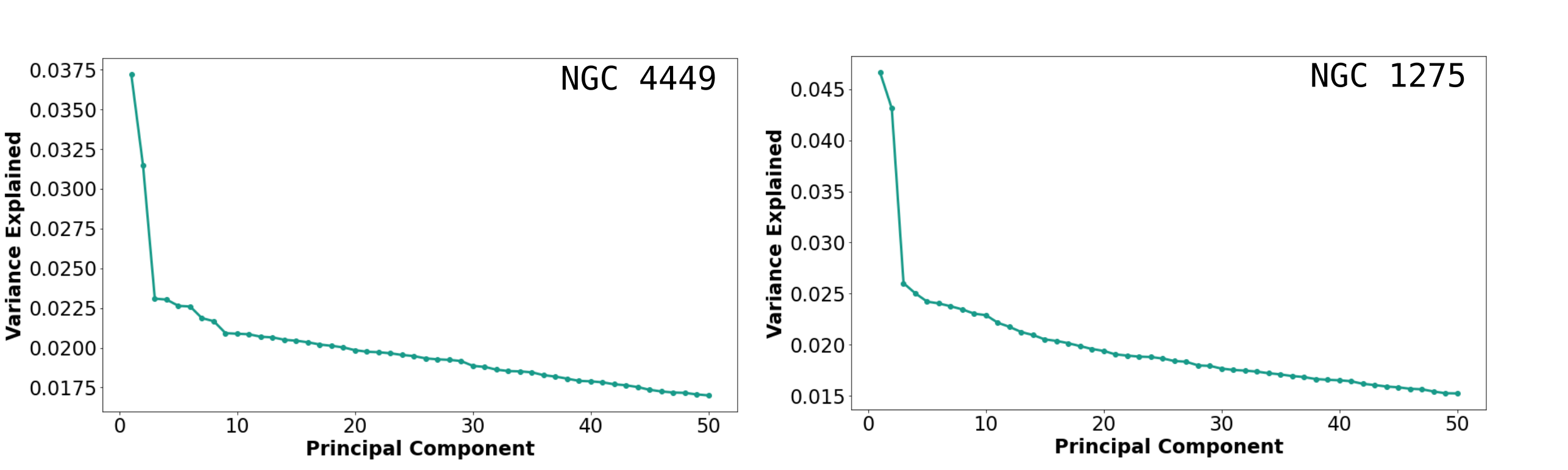}
    \caption{
       Scree plots for the PCA of NGC 4449 (left) and NGC 1275 (right). The scree plots show the explained variance ratio as a function of the principal component. Both scree plots show that the first two principal components explain the majority of variance in the observations; the results of a scree plot in addition to the visualization of principal components (figure \ref{fig:PCAcombined}) help determine the number of principal components required to accurately reproduce the background.
    }
    \label{fig:screePlots}
\end{figure*}

Interestingly, the scree plots reveal that the principal components explain a relatively low amount of the total variance in the background spectra; additionally, the eigenspectra show that only the first few components have emission/absorption features while the remaining components are purely noise. Taken together, this indicates that the assumption of linearity inherent to PCA may not be appropriate here. 
%However, in figures \ref{fig:exampleBKG} and \ref{fig:exampleBKG-NGC1275}, we show that the PCA reconstruction using only the first few components faithfully reconstructs the background spectra.
%We explore the possibility of a non-linear reconstruction in appendix \ref{app:nonlinear}. 
This indicates that the majority of the variance throughout the FOV is noise. However, since we see signal-related features over several eigenspectra, it may be the case that a non-linear reconstruction is necessary; this is beyond the scope of this work.

\subsection{Interpolation via an Artificial Neural Network}\label{sec:NN}
The next step is to train an artificial neural network to interpolate the \texttt{k}-dimensional coefficient vectors from the background pixels onto the source pixels.  
Thus, the network will take the pixel's x and y coordinates as input and return a k-dimensional vector in which each element corresponds to the pixel's $\alpha$ value where the $\alpha$ value is the coefficient of the eigenspectrum generated during the PCA.
This technique has been used extensively in the scientific literature (e.g., \citealt{llanas_constructive_2006}; \citealt{chen_rapid_1996}; \citealt{plaziac_image_1999}; \citealt{rigol_artificial_2001}). Indeed, this method is used frequently in geoscience to interpolate geophysical properties over maps and in computer science to interpolate network properties. To our knowledge, this is the first work in astronomy using this technique. We note that using standard interpolation techniques such as nearest neighbors, polynomial interpolation, or b-spline interpolation can result in numerous artifacts since the source regions can be large; this is a well-known issue with standard interpolation schemes. We explore the effects of different interpolation schemes in section \ref{sec:rec-bkg}.

Simple artificial neural networks are capable of learning an interpolation over sparsely sampled multi-dimensional space (i.e., \citealt{sivapragasam_simple_2010}; \citealt{rigol_artificial_2001}). Therefore, we use a simple neural network deployed in \texttt{tensorflow} (\citealt{abadi_tensorflow_2015}; \citealt{chollet_keras_2015}) consisting of 2 layers of 200 and 300 nodes respectively, each activated by the \texttt{tanh} function (\citealt{clevert_fast_2015}). We treat both structural parameters (such as the number of hidden layers and nodes) as hyperparameters in addition to the activation function, loss function, optimizer, and standard hyperparameters (i.e., the learning rate and learning rate decay factor). We optimized the hyperparameters of the network using \texttt{optuna} (\citealt{akiba_optuna_2019}).
We use the Huber loss function (\citealt{huber_robust_1964}), the \texttt{adam} optimizer (\citealt{kingma_adam_2017}), and a learning rate of $10^{-2}$. We apply a learning rate reducer that reduces the learning rate by a factor of 0.75 if the validation loss has not changed by more than a factor of 0.5 for five epochs. Additionally, we apply early stopping that will end training if the validation loss does not change for 10 epochs; the number of epochs is capped at 100.

We construct the training set by randomly selecting 99\% of the background pixels' PCA representation. The remaining percent is used to validate the training. Although this is a small percentage compared to the standard 10\%, we require a maximum number of spectra in the training set to ensure an appropriate spatial coverage. %We have tested the methodology using test sets containing 1\%, 2\%, 5\%, 10\%, and 20\% of the background pixels; we report no change in the loss as a function of the test set size.
Moreover, the goal in training this neural network is not to create a network that will generalize to other PCA coefficient maps but rather to learn how to interpolate for a specific set of maps that are unique to the observation being studied. While we are not concerned with generalizability, we do need to contend with overfitting, which is why we have a small holdout set. We use this holdout set to verify that the network is not overfitting the data. We note that this is standard in the relevant literature.

The artificial neural network outperforms standard interpolation techniques when the background covers a small fraction of the field of view and the masked out regions are large.
%large regions are masked out. 
However, standard interpolation methodologies can be used instead of the neural network for small masked regions. We have implemented \texttt{scipy.griddata} in the code as an alternative interpolation strategy using nearest neighbors or linear interpolation; these results are discussed in $\S$\ref{sec:rec-bkg}.
Once the coefficients are calculated for the masked-out spaxels, we can calculate a reconstructed background model by summing the principal component vectors multiplied by their corresponding coefficient and the mean background spectrum. We rescale these values in the same manner we initially scaled the data for the PCA.

\section{Results} \label{sec:results}
In this section, we apply the methodology discussed above to a galaxy and a galaxy cluster in order to showcase its efficacy in different test cases.

\subsection{NGC 4449}\label{sec:NGC4449}
In the following section, we discuss the application of the background construction algorithm to a SIGNALS galaxy \textbf{\href{http://cdsportal.u-strasbg.fr/?target=NGC\%204449}{\textit{\NGC4449}}}. \NGC4449 is an irregular, dwarf galaxy (M$_\star \approx 1.1\times 10^{11} \text{M}_\odot$) at a distance of approximately 3.8 Mpc (e.g. \citealt{annibali_cluster_2011}; \citealt{martinez-delgado_dwarfs_2012}; \citealt{sacchi_star_2018}). 
In figure \ref{fig:deepImages}, we show the SITELLE SN3 (6480 - 6850 \AA) deep image of \NGC4449. The image captures the active \hii{} regions leading to \NGC4449's extreme starburst ($0.47\text{ M}_\odot \text{yr}^{-1}$; \citealt{hunter_neutral_1999}; \citealt{sacchi_star_2018}), morphological features such as a northern bar, arcs and bubbles of ionized gas, and a persistent diffuse ionized gas permeating through the galaxy.
%Although we only explore the test case of a nearby galaxy, this methodology can be applied to any other SITELLE data (this is discussed further in section \ref{sec:appl}). 
Moreover, \NGC4449 is an interesting test case due to the omnipresence of this DIG throughout the galaxy, which poses a complication for proper background modeling. This complication arises because the theoretical modeling of the DIG is complex, and it is challenging to disentangle DIG emission from \hii{} region emission.

\begin{comment}
\begin{figure}
    \centering
    \includegraphics[width=0.485\textwidth]{Figures/NGC4449Deep.png}
    \caption{Deep image of the SN3 filter observation of \NGC4449 taken with SITELLE. The color scale was chosen to highlight the omnipresent DIG emission. Background and foreground objects appear as bright point sources in the image surrounding the galaxy. The cyan square represents the area over which the segmentation map was applied.}
    \label{fig:deep}
\end{figure}
\end{comment}

\subsubsection{Segmentation Maps}
The segmentation algorithm described in section \ref{sec:segAlgo} results in a mask of non-background pixels; figure \ref{fig:backgroundMapsCombined}a highlights emission regions while black pixels represent background spaxels. We note several structures exist in the segmentation map; primarily, the mauve central structure follows the emission of \NGC4449 nicely, including DIG regions. Several horizontal and vertical structures are marked as source regions but are, in actuality, simply observational artifacts of SITELLE owing to saturation spikes and the fact that the CCDs are combined in SITELLE. Since these spaxels are masked, they do not contribute to the background subspace. However, we are still able to obtain a background model for these spaxels after training the neural network for interpolation described in $\S$ \ref{sec:NN}. Additionally, point sources representing foreground or background active galactic nuclei (AGN) and stars are masked by the segmentation algorithm. Since we are only interested in modeling the background that will affect the flux measurements of NGC 4449, we crop the segmentation map only to include $400<x,y<1600$ corresponding to the cyan box in figure \ref{fig:deepImages}; the units are pixels. We apply this crop since, beyond this region, the cube does not contain visible emission from NGC 4449. Indeed, we also ran the method using a larger area and did not find any emission extending beyond the cyan box.

\begin{comment}
\begin{figure}
    \centering
    \includegraphics[width=0.485\textwidth]{Figures/BackgroundPixelMap_SN3.png}
    \caption{Segmentation map created from the deep image of \NGC4449. Each colored region represents a segment found by \texttt{photutils}. Colors are recycled, so each colored region does not necessarily represent a distinct segment. Note that the segmentation of each colored region has no physical representation and is an artifact of the algorithm. 
    }
    \label{fig:segmentationMap}
\end{figure}
\end{comment}
\subsubsection{Principal Component Analysis}
We present the first ten principal components of the background spectra in \NGC4449 in figure \ref{fig:PCAcombined}. Several notable features are present across the ensemble of components; most notably, the majority of sky emission lines are present in the mean spectrum, indicating that the mean spectrum is a good first-order approximation of the background. Additionally, the mean spectrum contains emission components typically associated with \hii{} regions or the DIG such as \nii{}\textlambda6548, \nii{}\textlambda6583, H\textalpha{}, \sii{}\textlambda6716, and \sii{}\textlambda6731. This is expected since there are undoubtedly interloping DIG emissions in the background spaxels due to the segmentation algorithm parameters chosen. 
For the purposes of this paper, it is good that the DIG is included in the background since we want to study \hii{} region emission and thus treat the DIG as a component of the background. If the goal was to calculate the total strong emission line flux (i.e., H\textalpha{}{}) in a given spaxel regardless of the emission region, then this would be inconsistent. It would be better to change the segmentation algorithm such that all regions included in DIG are not included in the background. We note that this is not an easy task and would likely require the inclusion of spectral information.
Alternatively, it is possible not to include the first principal component in the reconstructions; however, this assumes that the DIG emission is not present in the other components.

%The scree plot (see figure \ref{fig:Scree}) indicates that the majority of the variance is contained within the first ten components and that the first two components are the most important using the elbow method (e.g., \citealt{ferre_selection_1995}). 

The first component contains additional H\textalpha{}  emission and additional continuum. H\textalpha{}  emission is the primary emission line observed in the DIG, and the DIG is omnipresent in this galaxy, so it is reasonable that it represents an important component in the background spectra. Moreover, since the coefficients of the principal components can be negative, this can reflects the absence of H\textalpha{} emission in a considerable portion of background spectra. The scree plot reveals that this component explains nearly twice as much variance as any other component, thus indicating its importance (see figure \ref{fig:screePlots}). The second component shows negative flux in  \nii{}\textlambda6548, \nii{}\textlambda6583, H\textalpha{}, \sii{}\textlambda6716, and \sii{}\textlambda6731. This is a common signature of DIG emission (e.g., \citealt{vale_asari_importance_2021}). The remaining components, including those not shown here\footnote{We only plot the first ten principal components since all other components are primarily noise features, and their addition makes the graph unreadable.} primarily represents the noise signatures in the background spectra. Although the other components mainly represent noise, components two through eight have slight contributions near 6560 \AA; %therefore, we retain components one through eight in our reconstruction. 
%\textref{component 8 also corresponds with the knee of the scree plots (figure \ref{fig:screePlots}). Therefore, we retain the first eight principal components.}

We keep only the first three principal components since the scree plot (figure \ref{fig:screePlots}) shows a knee at that point.

\subsubsection{Reconstructed Backgrounds}\label{sec:rec-bkg}
We present here the reconstructed background regions. We show the coefficient maps of the first three principal components over the entire interpolation region, including both background and source pixels in figure \ref{fig:ML-interpolation}. We note that the first principal component coefficient peaks around the emission regions since it describes DIG emission; meanwhile, the second and third component coefficients are more homogeneously dispersed throughout the field (figure \ref{fig:ML-interpolation}). In figures \ref{fig:linear-interpolation} and \ref{fig:nearest-interpolation}, we show the results of using standard linear interpolation and nearest neighbor interpolation. Compared to the neural network reconstruction, the interpolated pixels show strong discontinuities and non-smooth behavior due to the nature of the interpolation strategies.

\begin{figure*}
    \centering
    \includegraphics[width=0.98\textwidth]{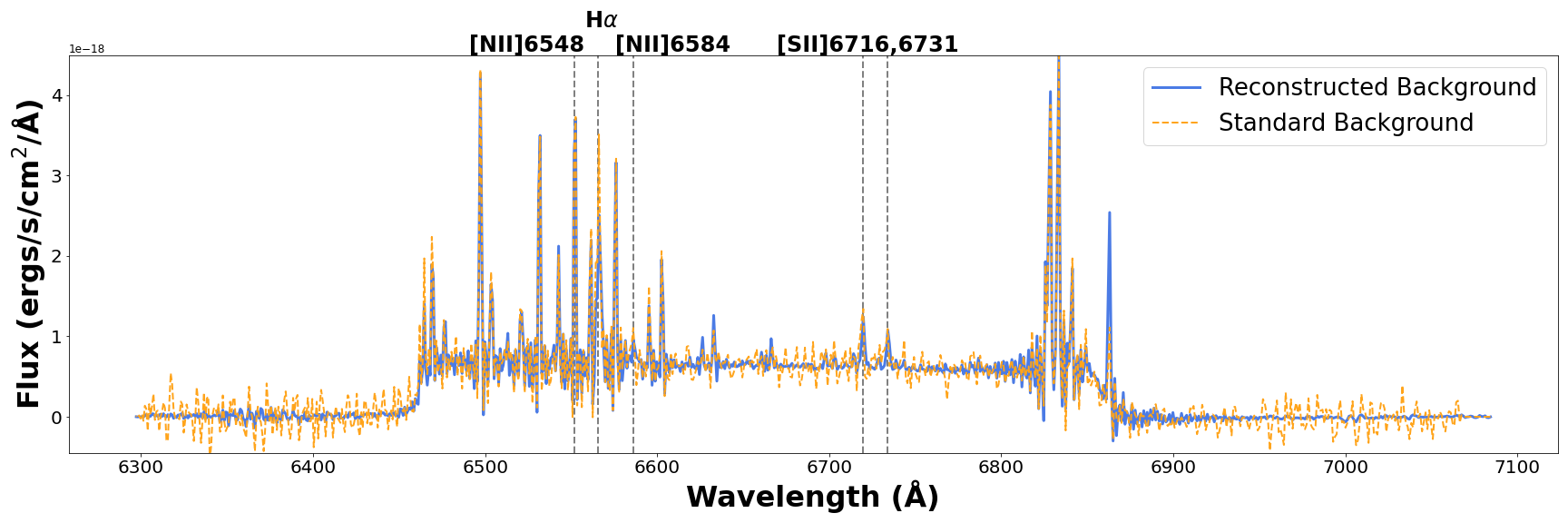}
    \caption{
    Here, we show the standard background versus the PCA reconstructed background for the same background region of NGC4449. Vertical, gray, and dashed lines indicate strong emission lines. In addition to reducing the noise in the background spectrum, the reconstructed method changes the background flux, slightly affecting the measured fluxes of strong emission lines. The emission lines present in the background represent DIG emission.
    }
    \label{fig:exampleBKG}
\end{figure*}

In figure \ref{fig:exampleBKG}, 
%we show the mean and reconstructed background spectrum using the first three principal components for a randomly selected source pixel.
we show the standard background versus the PCA reconstructed background for the same region.
The graphic demonstrates the importance of using a local reconstructed background by highlighting the changes in line amplitudes in spectral regions where strong emission lines (e.g., H\textalpha{}) are present.

\subsection{NGC 1275}\label{sec:NGC1275}
In this section, we apply our methods to the SITELLE observations of the well-studied galaxy cluster \textbf{\href{http://cdsportal.u-strasbg.fr/?target=NGC\%201275}{\textit{\NGC{}1275}}}. NGC 1275 is the brightest center galaxy in the Perseus cluster; it hosts a wide range of multi-wavelength astrophysical phenomena (\citealt{fabian_relationship_2003}; \citealt{fabian_wide_2011}; \citealt{krabbe_near_2000}; \citealt{hitomi_collaboration_quiescent_2016}; \citealt{gendron-marsolais_revealing_2018}; \citealt{vigneron_high-spectral-resolution_2024}). In the optical bandpass, NGC 1275 hosts a large (several dozens of kiloparsecs) and asymmetric emission-line nebula. Unlike the previous test case of \NGC 4449, \NGC 1275 does not exhibit DIG that needs to be disentangled from the nebular emission; additionally, the background is expected to be relatively stable over the field (\citealt{gendron-marsolais_revealing_2018}; \citealt{vigneron_high-spectral-resolution_2024}).

\subsubsection{Segmentation Map}
The segmentation map is shown in figure \ref{fig:backgroundMapsCombined}b. We used the same segmentation algorithm to create this map but changed the $\sigma$-threshold to 0.5 to find a better contrast between the background and the nebula. This value was experimentally determined. 
Similar to the segmentation map of NGC 4449, background pixels are shown as black pixels. Also, all point sources are masked and thus show up as black pixels. The deep image is cropped to include only $800<x<1600$ and $200<y<1200$ since the rest of the observation does not contain nebular emission; the units are in pixels. 
Several sources of note are contained within the mask; the large circular regions, excluding that in teal centered at approximately pixel (400, 250), indicate emission from background and foreground galaxies, which should be excluded from the background model. Additionally, the nebula itself, which is made up of several segments, including the central teal segment, is included in the mask. The horizontal lines near the bottom right are saturation spikes in the SITELLE data.

\subsubsection{Principal Component Analysis}
In figure \ref{fig:PCAcombined}, we show the first ten principal components ordered by explained variance (importance) for NGC 1275. 
Similar to the PCA results of NGC 4449, the mean emission contains the sky-lines and is a good first-order approximation of the background. Unlike NGC 4449, the principal components do not contain emission lines typical of strong emission lines. In the case of NGC 1275, there is no diffuse nebular emission (the DIG in NGC 4449), so the segmentation algorithm completely blocks out all emission associated with the nebula. Components 1 and 2 show that there is slight variation in the sky-lines across the observation and that the majority of the variance occurs near the \sii{}-doublet and on the edges of the transmission region of SN3 (at $\approx$ 6350 \AA  and $\approx$ 6750 \AA). We note that the overall variance explained is low; however, the components after component 2 clearly show noise. Thus, we only use the first two components in our background reconstructions.

\begin{comment}
\begin{figure}
    \centering
    \includegraphics[width=0.495\textwidth]{Figures/PCA_components_normalized_SN3_NGC1275.png}
    \caption{This graphic shows the first ten principal components that describe the background emission in NGC 1275. We have normalized the spectra by their maximum value. We do not include higher-order components since they contain only noise.}
    \label{fig:pca-NGC1275}
\end{figure}
\end{comment}

\subsubsection{Reconstructed Backgrounds}
In figure \ref{fig:exampleBKG}, we show the standard background versus the PCA reconstructed background for the same background region. 
Due to the higher redshift of NGC 1275 compared to that of NGC 4449, the positions of the strong emission lines are shifted a considerable amount to longer wavelengths. This moves the H\textalpha{} complex out of the forest of skylines between 650 and 660 nanometers, allowing for easier measurements. However, the \sii{}-doublet is shifted into the spectral bandpass occupied by skylines  around 685 nanometers. Because of this, in order to model the \sii{}-doublet, it is crucial to have an accurate background model. 

We note that the noise level in the reconstructed background is considerably lower than the noise in the standard background for both observations. This effect is most evident outside of the transmission region (for example, between 6300 and 6450 \AA). This is due to the fact that we disregard high-order eigenspectra that capture this noise. We choose to do this to obtain a background model containing as little noise as possible in order not to inject noise into the background subtracted spectrum we eventually use for fitting to reduce the overall noise level. This choice comes with a potential bias to the signal. Users can choose between this spectral bias and the noise level by opting to include more or less eigenspectra.
%However, users can choose to include more noise in the model by keeping more eigenspectra in the reconstruction (see appendix XXX for an example).

%The graphic demonstrates the importance of using a local reconstructed background by highlighting the changes in line amplitudes in spectral regions where strong emission lines (e.g., H$\alpha$) are present. 
%For example, at the location of H$\alpha$, the amplitude of the mean emission of that wavelength is higher than the reconstructed background but considerably lower than the reconstructed source. This implies that if one uses the mean emission spectrum, the background contribution of H$\alpha$ will be underestimated, and the final H$\alpha$ flux calculated for the spaxel will be overestimated for this particular pixel.

\begin{figure*}
    \centering
    \includegraphics[width=0.98\textwidth]{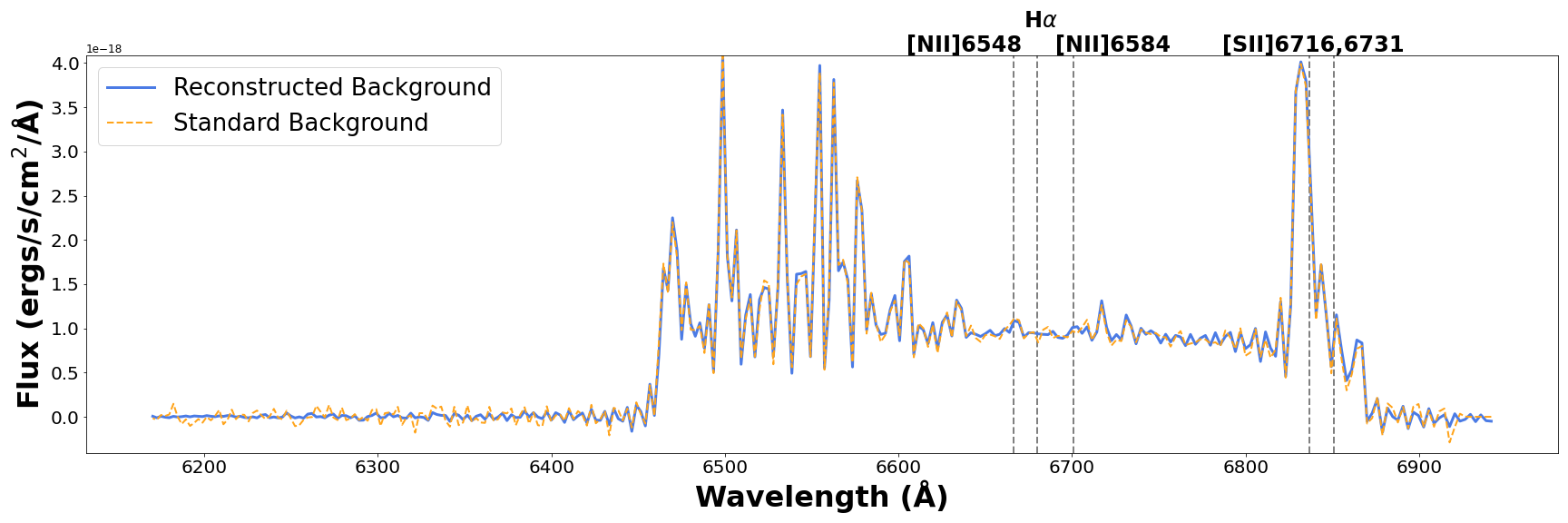}
    \caption{
    Here, we show the standard background versus the PCA reconstructed background for the same background region comprising approximately 100 pixels of NGC 1275. Vertical, gray, and dashed lines indicate strong emission lines redshifted to Perseus' global velocity. %In addition to reducing the noise in the background spectrum, the reconstructed method changes the background flux, affecting the measured fluxes of strong emission lines.
    }
    \label{fig:exampleBKG-NGC1275}
\end{figure*}

\section{Discussion}\label{sec:discussion}

In the previous section, we presented the results of our methodology. We demonstrated how the SITELLE deep image can be used to construct a segmentation map, created a principal component decomposition of the spectra belonging to the background, and presented the reconstructed backgrounds using our neural network trained on the PCA coefficients of the background spaxels. In the following section, we compare extracted flux maps using our methodology to the standard methodology in NGC 4449 and NGC 1275. We conclude with a discussion of other uses for this background methodology.

\subsection{Advantages over Standard Methodologies}

The methodology proposed here does not assume any homogeneity in the background spectra. If the background does not vary in a given cube, then the coefficients for the eigenspectra will be near zero. Any deviations from homogeneity will be encoded in spatial changes in these coefficients. This method works equally well for systems with complicated emission region complexes and systems with simple emission regions.

This section compares how the different background modeling methods affect the final calculated flux of strong emission lines in source regions. We fit our cubes using \texttt{LUCI} after subtracting a single global background using the standard methods detailed in section \ref{sec:methods} and comparing the combined H\textalpha{} and \nii{}-doublet amplitude to that calculated using our interpolated background scheme. We present an example background spectra using the two methods in figure \ref{fig:exampleBKG} and in figure \ref{fig:exampleBKG-NGC1275} -- see $\S$ \ref{sec:rec-bkg} for a detailed comparison. 
%Although we only detail the H$\alpha$ emission line, these results hold for the other lines present in the spectra. 

We fit the unbinned data with a \texttt{sinc} function tieing the five emission lines in velocity space; the five emission lines are \nii{}\textlambda6548, \nii{}\textlambda6583, H\textalpha{}, \sii{}\textlambda6716, and \sii{}\textlambda6731. We fit the cube twice separately: once using the novel background method and again using the standard background methodology. For the standard methodology, we selected a background region sufficiently far from NGC 4449\footnote{We used a circular region centered at (\textit{12:27:53.2,+44:07:59.2}) and a radius of 19 arcseconds} (and in NGC 1275\footnote{We used a circular region centered at (\textit{3:19:42.5,+41:32:01.8}) and a radius of 10 arcseconds}) such that it does not contain emissions from the galaxy. 

\begin{figure*}
    \centering
    \includegraphics[width=0.99\textwidth]{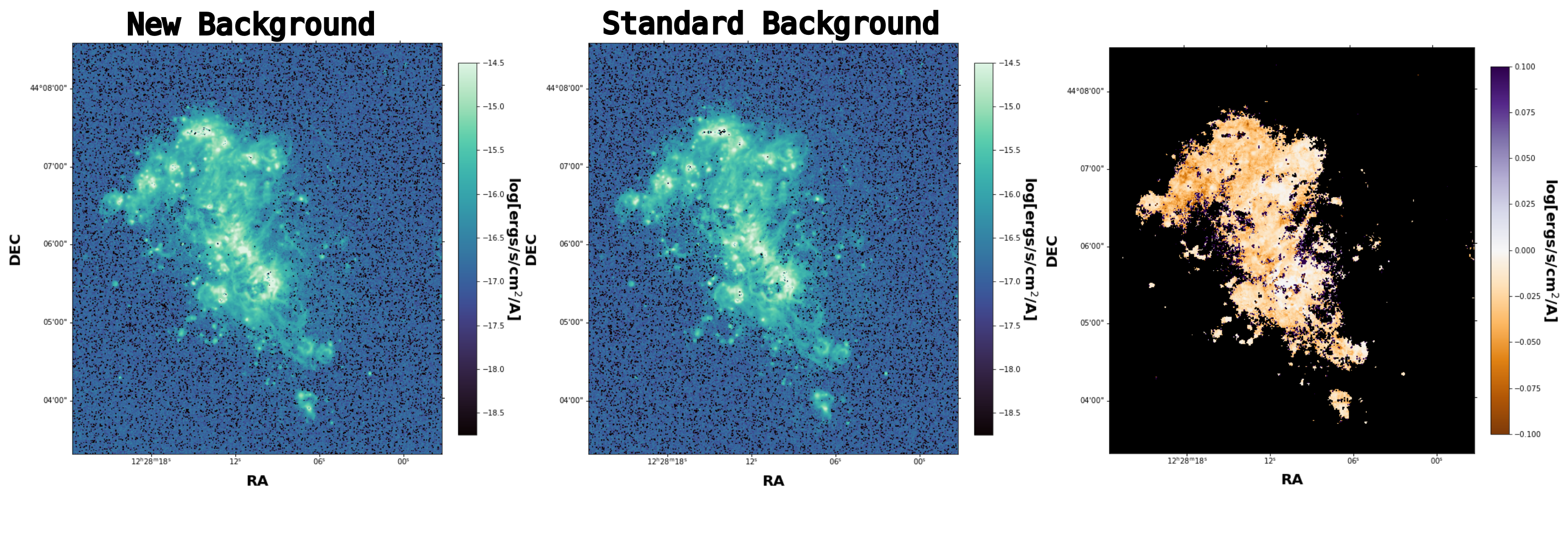}
    \caption{Combined amplitude fit (H\textalpha{} + \nii{}-doublet) for NGC 4449. On the left, we have the fit using the background method presented in this article. In the middle, we show the fit using the standard background methodology. Finally, on the right, we show the difference map between these two. Pixels with a combined log amplitude under -17.5 are masked to highlight the differences.}
    \label{fig:NGC4449-fit}
\end{figure*}

In figure \ref{fig:NGC4449-fit}, we show the combined H\textalpha{} and \nii{}-doublet amplitude map for the new background model (left), the standard method (center), and the difference between the two maps (right; the difference map is new background model map minus the standard background map) for NGC 4449. The difference map reveals that, in the galaxy's central regions, the standard background methodology overestimates the flux; however, we can see a stark change in the fits in the outer regions. The purple pixels indicate that more flux (using the amplitude as a proxy) is recovered using the new methodology. We expect this since we are now correctly modeling the background in the DIG.

\begin{figure*}
    \centering
    \includegraphics[width=0.99\textwidth]{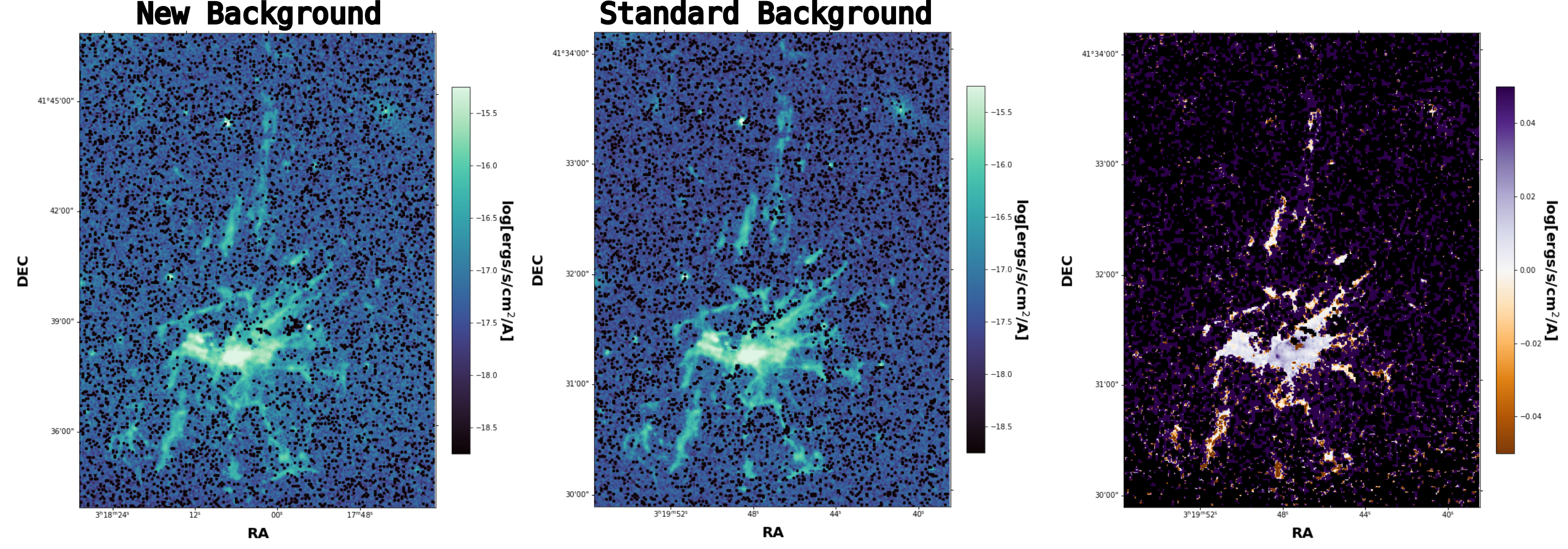}
    \caption{From left to right: Combined H\textalpha{} and \nii{}-complex amplitude map using the background method presented in $\S$ \ref{sec:methods}, Combined H\textalpha{} and \nii{}-complex amplitude map using the standard background method, and the difference map between the two maps generated using different background models for NGC 1275.}
    \label{fig:NGC1275-FitComparison}
\end{figure*}

In figure \ref{fig:NGC1275-FitComparison}, we show the amplitude of the combined H\textalpha{} and \nii{}-doublet fit using the new background technique (left), the standard background method (center), and the difference between the two (right) for NGC 1275. The standard background region was taken from near the nebula, but, importantly, it does not include any nebular emission or point sources. Unlike the results for NGC 4449, the difference map reveals that using the new background method, we recover slightly lower amplitudes on the perimeter of the nebula. This indicates that the standard background methodology underestimates the background emission in these areas. For both objects, we verified that the change in background modeling did not affect the velocity or velocity dispersion values.

\subsection{Potential Modifications}
While we have highlighted the use of this methodology on \NGC4449, it can be extended to any other galaxy in the SIGNALS catalog. Moreover, it can be applied to other SITELLE observations. However, the interpretation of the principal components will likely change as the primary background contaminant may change (i.e., in galaxy clusters, the primary contaminant is expected to be emission from the stellar continuum rather than DIG). This method could also be adjusted to model the stellar continuum by modifying the background segmentation algorithm to only place spaxels containing stellar continuum emission in the background space. While out of the scope of this article, this topic will be explored in further studies. Furthermore, the algorithm is not SITELLE specific but can be used for any IFU data with sufficient spatial resolution to capture background emission in the spaxels (see $\S$ \ref{sec:generalization}). 

Although we run the background detection algorithm on the deep image, in some cases, such as extreme stellar contamination, which muddies the deep image, using a flux or line-amplitude map of the strongest emission line is advantageous. In doing so, we ensure that only the line-emitting regions are masked. This application is explored in other works (i.e., Hlavacek-Larrondo in prep.; Rhea et al. in prep.). In addition to changing the image used to detect background regions, this methodology also works with different segmentation algorithms (i.e., thresholding algorithms).

In the two examples showcased in section \ref{sec:results}, the neural network interpolations yielded smooth reconstructions of the coefficient fields. However, it is possible that users may want less smooth interpolations in the case where high-frequency features are present in the coefficient maps. Inherently, neural fields, such as the one developed in this work, return smooth representations. In order to capture high-frequency (i.e. non-smooth) features in the interpolation, it is possible to add Fourier features to the input vector (\citealt{tancik_fourier_2020}); this will be explored in future works.

\subsection{Generalization to Other Instruments}\label{sec:generalization}
While we showcased this methodology using data from SITELLE, it can be readily ported to any IFU-style data such as MUSE (\citealt{bacon_muse_2017}). Since this method does not rely on certain emission lines to constrain the background emission, it is not limited to a given spectral range. In the next subsections, we explore uses in other wavelengths.

\subsubsection{Optical}
Due to the nature of its design, SITELLE is built to capture primarily emission lines (see \citealt{drissen_sitelle_2019} for details). Our methodology will work equally well for absorption lines. Therefore, it can be used for galactic and solar studies using instruments such as GMOS (\citealt{allingtonsmith_integral_2002}),  MUSE (\citealt{bacon_muse_2017}), and MEGARA (\citealt{gil_de_paz_megara_2012}). For example, instead of using dedicated background fibers that implicitly assume a homogenous background, such as those in the GMOS instrument, the background can be directly sampled and modeled from the science fibers using our methodology.
Moreover, optical IFU-like instruments, such as the GH{\textalpha{}}FaS instrument on the William Herschel Telescope (\citealt{hernandez_gh_2008}), can also be used. 

\subsubsection{Infrared}
With the advent of the James Webb Space Telescope (JWST) we have access to an IFU in the near-infrared (NIR) and mid-infrared (MIR) bandpasses (e.g., \citealt{gardner_james_2006}; \citealt{rieke_mid-infrared_2015}; \citealt{gordon_mid-infrared_2015}; \citealt{pontoppidan_jwst_2022}; \citealt{jakobsen_near-infrared_2022}; \citealt{boker_near-infrared_2022}). Methods for modeling the background in the NIR will suffer from the same complications as optical IFUs; therefore, it is important to have a robust methodology for handling the background. Since our methodology is wavelength-independent, it can be used for JWST NIRSpec IFU and MIRI IFU observations.   This methodology will complement the strategies outlined by the JWST team (e.g., \citealt{boker_near-infrared_2022}) to reduce background contamination in IFU measurements. More specifically, this method can help achieve the primary science goals.
For instance, developing an accurate background model is crucial in studying faint distant galaxies. In nearby galaxies, the primary component of the background emission is expected to be emission from the host galaxy's stellar component and, therefore, requires dedicated modeling; our methodology allows this without assuming an underlying stellar population for the host galaxy. For near-earth objects, a complete background model is required to obtain accurate chemical abundances that will aid our understanding of our solar system's evolution and chemical makeup.

\subsubsection{X-ray}
The methodology outlined in this article does not only apply to IFU data but to any detector that records a photon's energy at each pixel; therefore, it is a perfect candidate for the ACIS instruments on the Chandra X-ray Observatory (e.g. \citealt{weisskopf_chandra_2000}; \citealt{weisskopf_overview_2002}; \citealt{garmire_advanced_2003}). The ACIS instrument includes two sets of CCD detectors. When a photon falls on the detector, the CCD records the position of the photon (i.e., the pixel), the time of the event, and the measured energy of the incident photon. For diffuse targets such as galaxy clusters, the background can play a crucial role in measurements (e.g., \citealt{miller_outer_2012}; \citealt{george_comprehensive_2014}). Several methods have been specifically developed to model the X-ray background, but they  depend on a concrete physical understanding of the mechanisms influencing background emission (e.g., \citealt{bartalucci_chandra_2014}; \citealt{markevitch_chandra_2003}). Contrarily, the methodology outlined here does not assume any physical models. Again, this method does not assume homogeneity of the background as many currently implemented methodologies for Chandra do (e.g., \citealt{vikhlinin_chandra_2006}; \citealt{cavagnolo_intracluster_2009}; \citealt{sun_chandra_2009}).

While this methodology can be applied to existing X-ray observatories, it is uniquely adapted to the requirements of future missions such as XRISM and Athena (\citealt{xrism_science_team_science_2020}; \citealt{simionescu_constraining_2019}; \citealt{nandra_hot_2013}; \citealt{barret_athena_2018}). The IFUs on these missions will not only map out the complex morphologies of galaxies and galaxy clusters but also the spectra in the background regions. Due to the enhanced spectral resolution of the detectors on XRISM and Athena, the background spectra will be considerably more complicated than what is currently considered and modeled. Therefore, the methodology outlined in this paper can serve as an alternative to physical modeling while achieving a faithful background spectral model.

%\subsection{Including Uncertainties}
%\carter{MDN}

%\carter{Other IFUs}

%\carter{Time Variability}

%\carter{Webb -- wide band image interpolation}

%\carter{Potential Uncertainties -- MDN}

%\subsection{Implementation in \texttt{LUCI}}
%\carter{Basic description of the standard \texttt{LUCI} implementation including how the algorithm has been wrapped to work with the standard fitting. Include descriptions and links to example notebooks.}

\section{Conclusions}
We present a novel strategy for modeling the background emission (both the sky and contaminant emission) for IFU-style instruments. This methodology uses a combination of image segmentation algorithms and PCA to model the background in regions where no source emission is present. We then developed an artificial neural network to interpolate the model over masked source regions.

This methodology is applied to a nearby irregular dwarf galaxy, NGC 4449, observed as part of the SIGNALS collaboration. We demonstrate the importance of using our background strategy on this target; since NGC 4449 is highly irregular and contains overlapping \hii{} regions and widespread diffusion ionized gas, other methodologies are not appropriate for background modeling. By comparing the recovered fluxes as computed by \texttt{LUCI} using different background methods, we demonstrate how our methodology allows us to obtain flux estimates for this system. 
We also repeat the experiment for the galaxy cluster NGC 1275, showing how the algorithm performs on a different type of object.
Finally, we consider how the algorithm, or rather certain aspects of the algorithm, can be modified or substituted to apply it to other related applications.
The algorithm has been implemented in the \texttt{LUCI}. An example jupyter notebook can be found at \url{https://github.com/crhea93/LUCI/blob/main/Examples/BackgroundAutomatic.ipynb}.

\section*{Acknowledgements}

The authors would like to thank the Canada-France-Hawaii Telescope (CFHT) which is operated by the National Research Council (NRC) of Canada, the Institut National des Sciences de l'Univers of the Centre National de la Recherche Scientifique (CNRS) of France, and the University of Hawaii. The observations at the CFHT were performed with care and respect from the summit of Maunakea which is a significant cultural and historic site.
C.L. R. acknowledges financial support from the physics department of the Universit\'e de Montr\'eal, the MITACS scholarship program, and the IVADO doctoral excellence scholarship.
J. H.-L. acknowledges support from NSERC via the Discovery grant program, as well as the Canada Research Chair program.
M. P. acknowledges financial support from the European Union’s Horizon 2020 research and innovation program under the Marie Skłodowska-Curie grant agreement No. 896248.
LRN is grateful to the National Science foundation NSF - 2109124 and the Natural Sciences and Engineering Research Council of Canada NSERC - RGPIN-2023-03487 for their support.

We used the following software: \texttt{tensorflow} (\citealt{abdi_principal_2010}, \texttt{keras} (\citealt{chollet_keras_2015}), \texttt{python} (\citealt{van_rossum_python_2009}, \texttt{scipy} (\citealt{virtanen_scipy_2020}), \texttt{matplotlib} (\citealt{hunter_matplotlib_2007}, \texttt{scikit-learn} (\citealt{pedregosa_scikit-learn_2011}), \texttt{optuna} (\citealt{akiba_optuna_2019}, \texttt{photutils} \citealt{bradley_astropyphotutils_2023}, \texttt{astropy} (\citealt{robitaille_astropy_2013}), \texttt{LUCI} (\citealt{rhea_luci_2021}), \texttt{ds9} (\citealt{joye_new_2003}).
We note that the versions of each software can be found on the \href{https://github.com/crhea93/LUCI}{\faicon{github}\texttt{LUCI}} github page in the \textit{requirements.txt} file.
%%%%%%%%%%%%%%%%%%%%%%%%%%%%%%%%%%%%%%%%%%%%%%%%%%
\section*{Data Availability}

All methods used in this paper are available at 
 \href{https://github.com/crhea93/LUCI}{\faicon{github} crhea93:LUCI}. The data can be accessed at the \href{Canadian Astronomical Data Centre}{https://www.cadc-ccda.hia-ihaffecteda.nrc-cnrc.gc.ca/}.



%\bibliographystyle{mnras}
% Beware for some reason using rasti.bst does not troncate long author lists... reverting to mnras.bst does, but might not be legit
\bibliographystyle{rasti}
\bibliography{SitelleBackground} % if your bibtex file is called example.bib

%%%%%%%%%%%%%%%%%%%%%%%%%%%%%%%%%%%%%%%%%%%%%%%%%%

%%%%%%%%%%%%%%%%% APPENDICES %%%%%%%%%%%%%%%%%%%%%

\appendix
\section{Interpolation Schemes}
\subsection{Neural Network Interpolation}

This figure is cited in the text but placed here for readability. 
\begin{figure*}
    \begin{subfigure}{0.33\textwidth}
        \includegraphics[width=\textwidth]{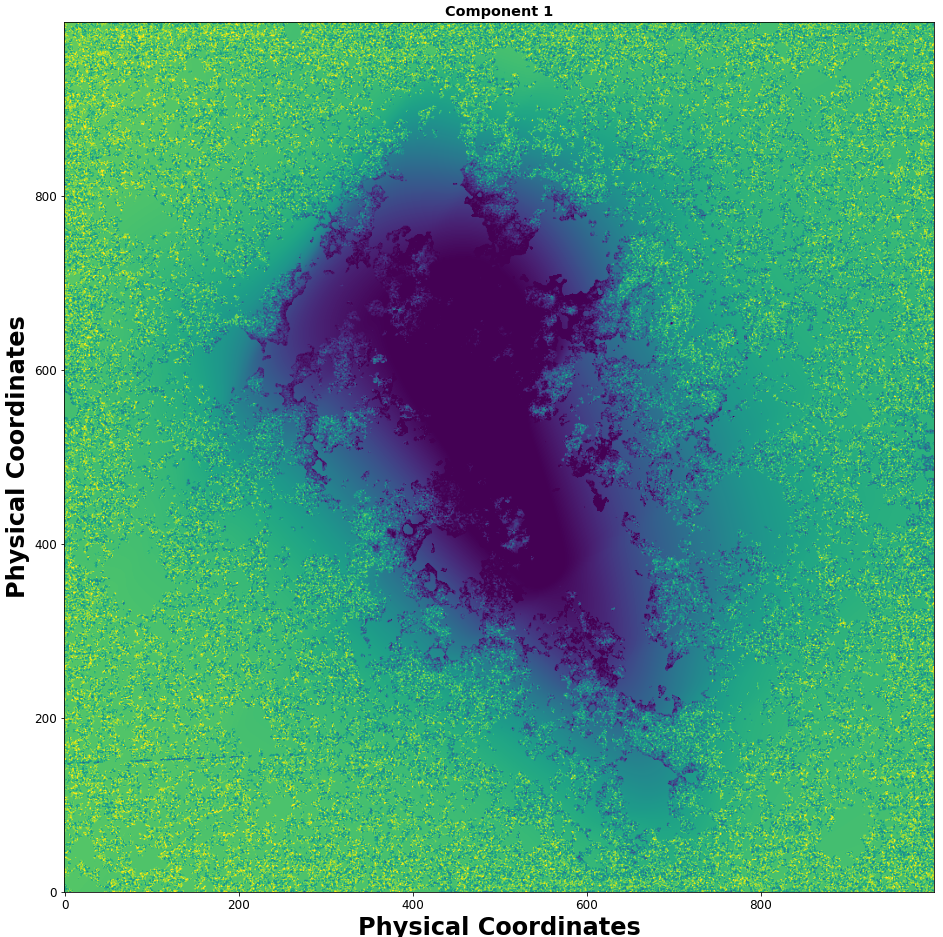}
        \caption{Principal Component 1}
        \hfill
    \end{subfigure}
        \begin{subfigure}{0.33\textwidth}
        \includegraphics[width=\textwidth]{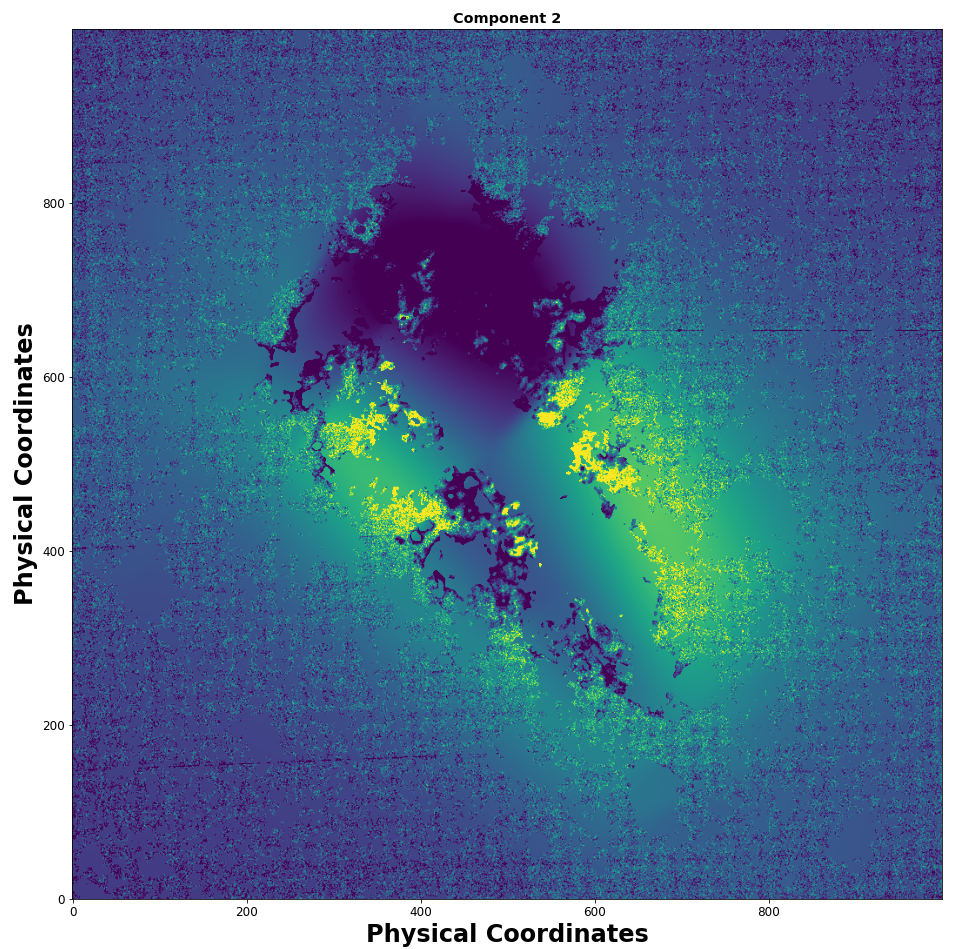}
        \caption{Principal Component 2}
        \hfill
    \end{subfigure}
        \begin{subfigure}{0.33\textwidth}
        \includegraphics[width=\textwidth]{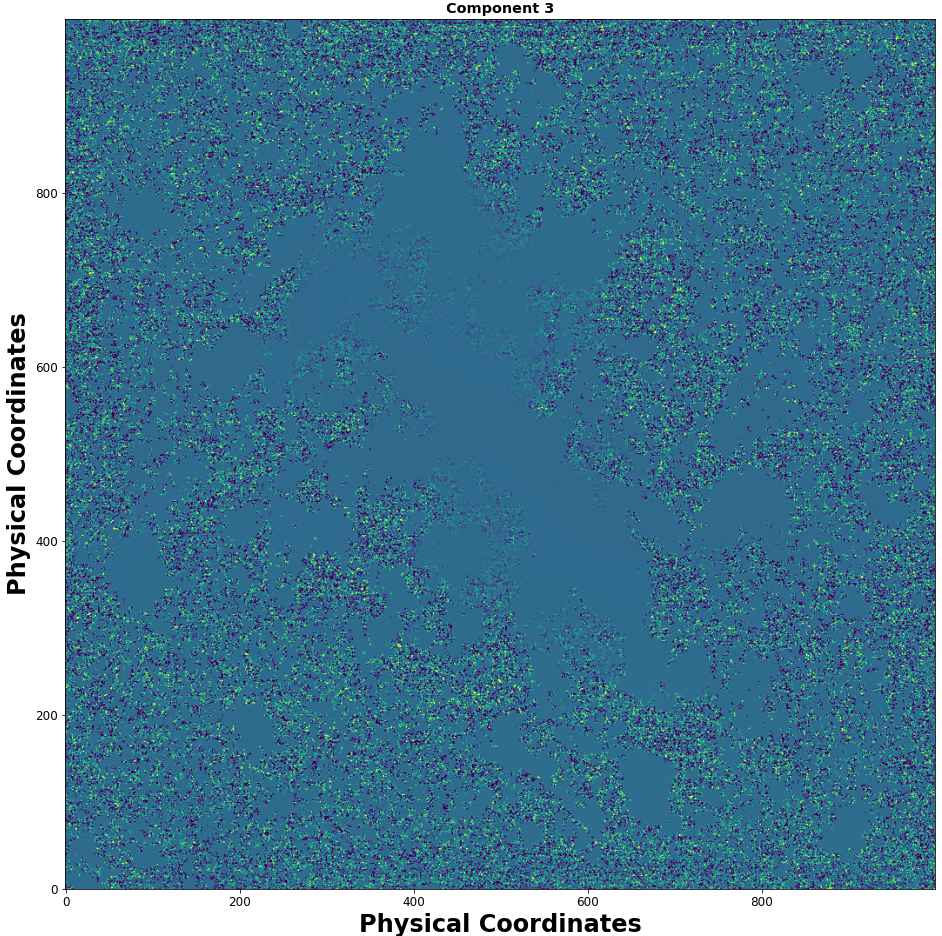}
        \caption{Principal Component 3}
        \hfill
    \end{subfigure}

    \caption{Interpolation of principal component coefficients using neural network interpolation. The coefficients have been scaled to unity; yellow implies a coefficient value near 1, and dark blue implies a value near 0. Although the background and source pixels are shown in the same color, they are distinguishable by the smoothness of the coefficients in the source regions. For example, regions of background pixels are remarkable in the presence of noise in the coefficient values.
    We note that these coefficients are unitless.
    }
    \label{fig:ML-interpolation}
\end{figure*}

\subsection{Linear Interpolation}
\begin{figure*}
    \begin{subfigure}{0.33\textwidth}
        \includegraphics[width=\textwidth]{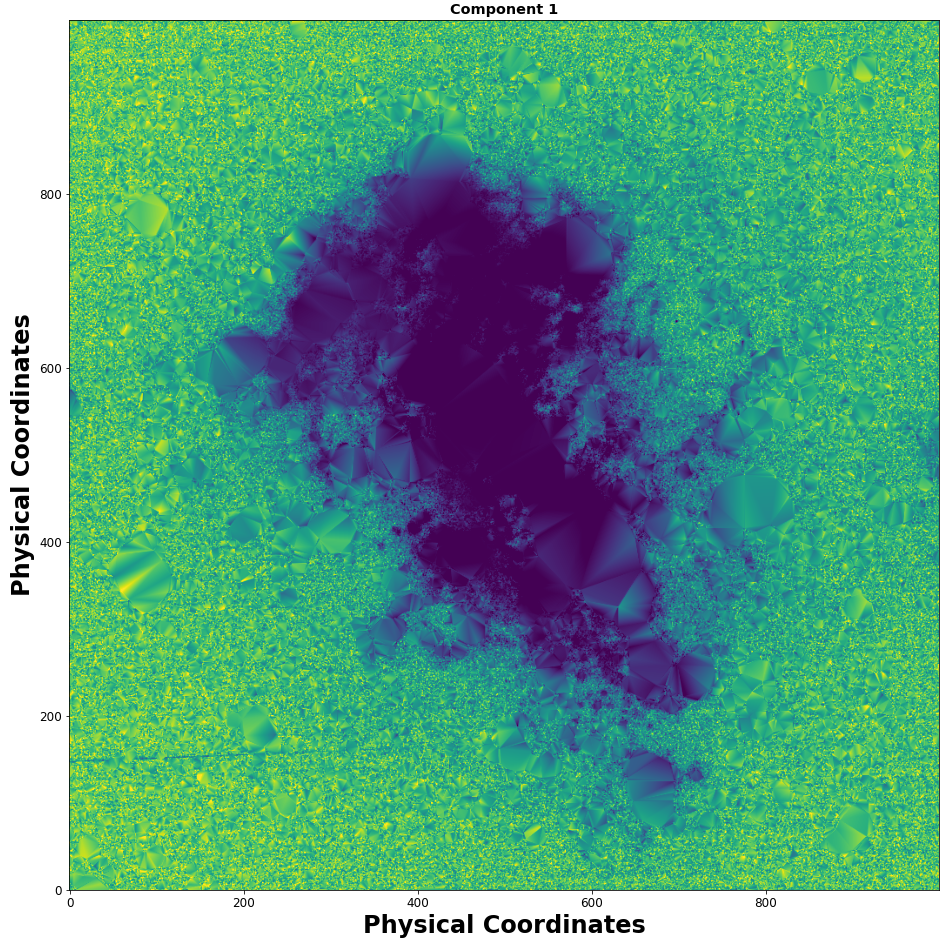}
        \caption{Principal Component 1}
        \hfill
    \end{subfigure}
    \begin{subfigure}{0.33\textwidth}
        \includegraphics[width=\textwidth]{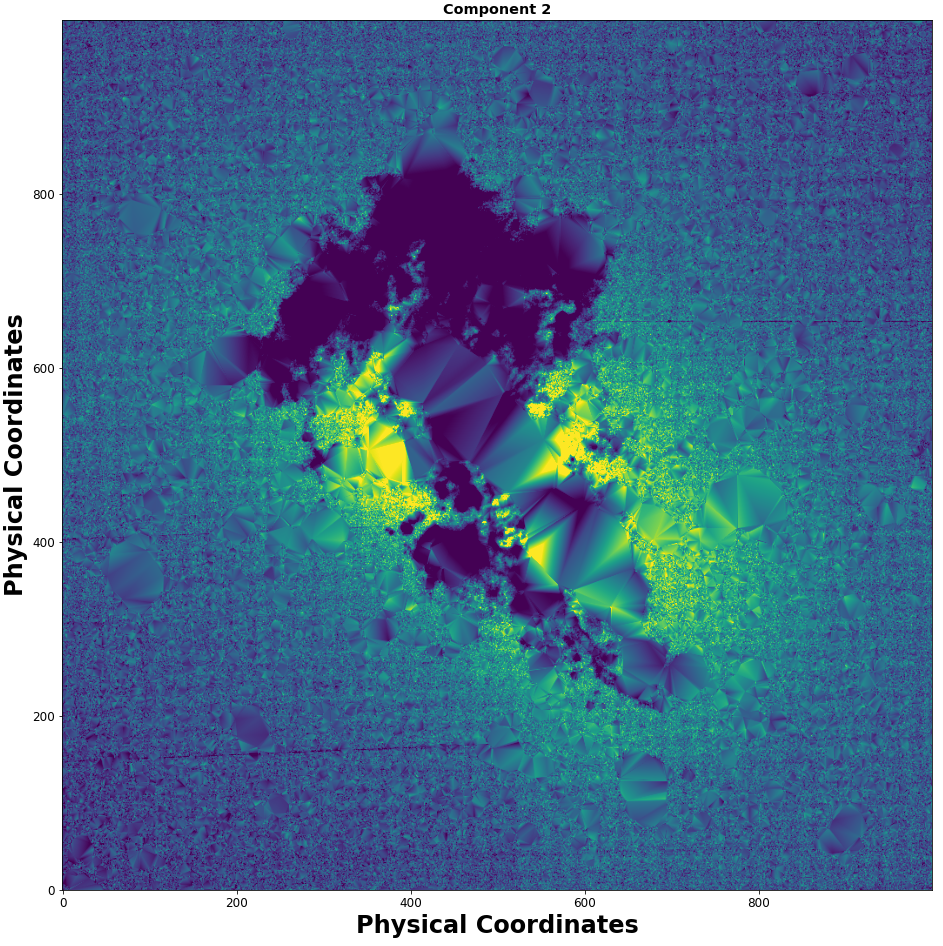}
        \caption{Principal Component 2}
        \hfill
    \end{subfigure}
        \begin{subfigure}{0.33\textwidth}
        \includegraphics[width=\textwidth]{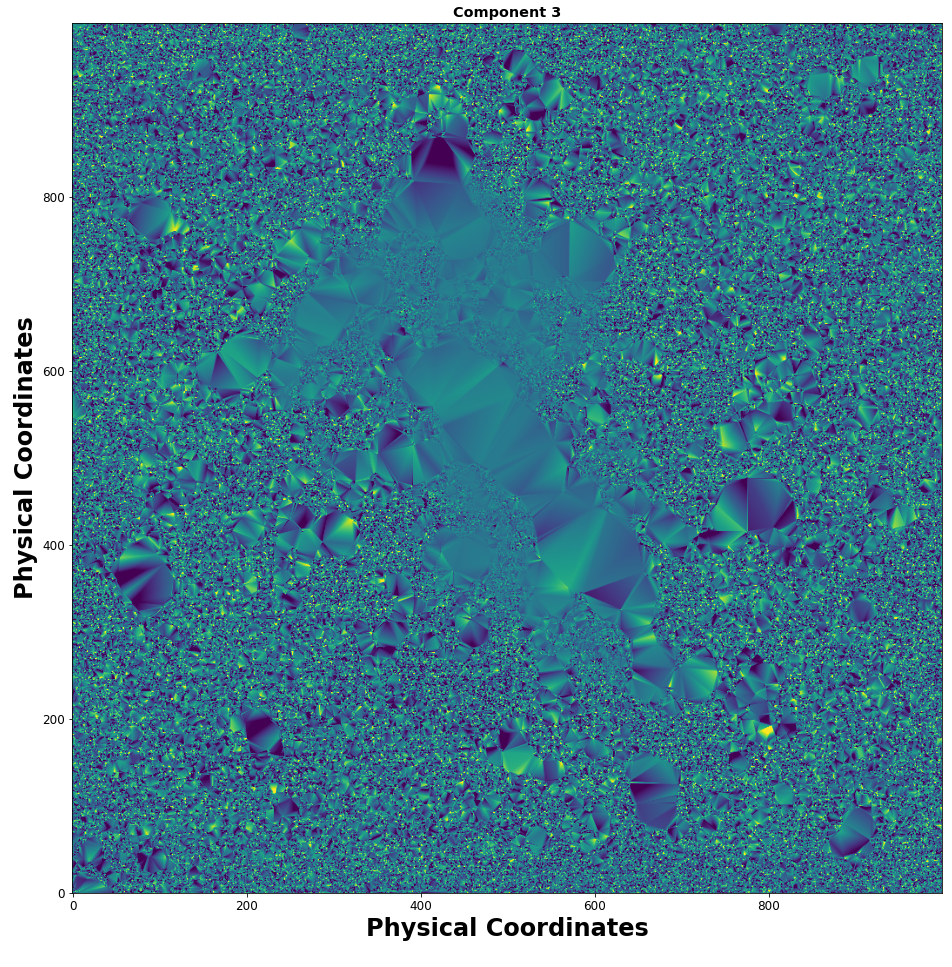}
        \caption{Principal Component 3}
        \hfill
    \end{subfigure}

    \caption{Interpolation of principal component coefficients using linear interpolation. The coefficients have been scaled to unity; yellow implies a coefficient value near 1, and dark blue implies a value near 0. Although the background and source pixels are shown in the same color, they are distinguishable by the smoothness of the coefficients in the source regions. For example, regions of background pixels are remarkable in the presence of noise in the coefficient values.
    We note that these coefficients are unitless.}
    \label{fig:linear-interpolation}
\end{figure*}

The plots shown in \ref{fig:linear-interpolation} demonstrate the results of using linear interpolation as implemented in \texttt{scipy.griddata(method='linear')}. By design, the interpolation has strong discontinuities in the interpolated regions which can lead to inaccurate background reconstructions.

\subsection{Nearest Neighbor Interpolation}
\begin{figure*}
    \begin{subfigure}{0.33\textwidth}
        \includegraphics[width=\textwidth]{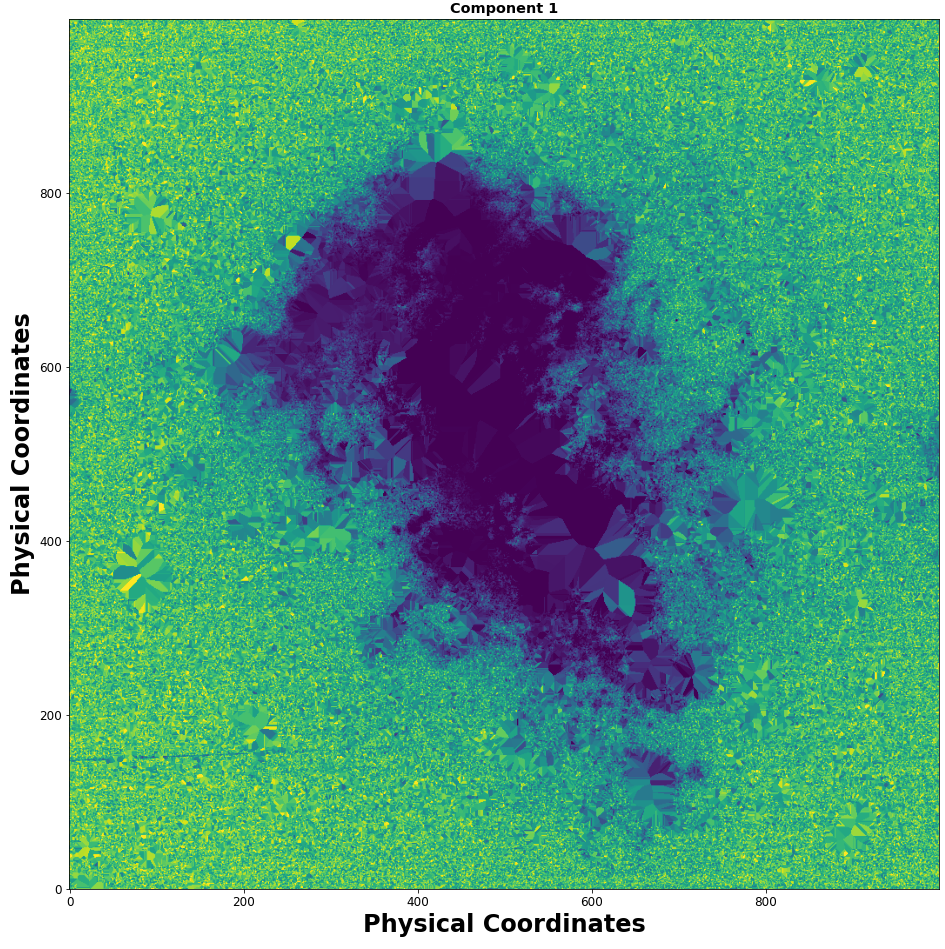}
        \caption{Principal Component 1}
        \hfill
    \end{subfigure}
    \begin{subfigure}{0.33\textwidth}
        \includegraphics[width=\textwidth]{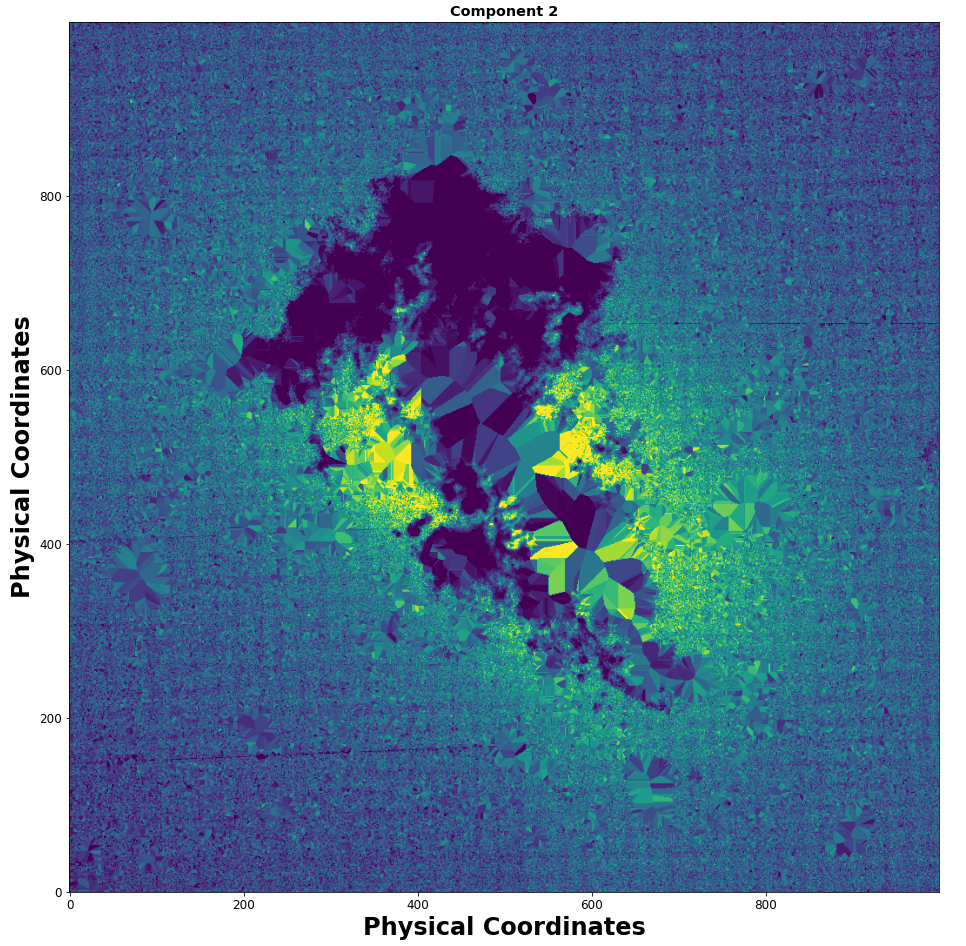}
        \caption{Principal Component 2}
        \hfill
    \end{subfigure}
        \begin{subfigure}{0.33\textwidth}
        \includegraphics[width=\textwidth]{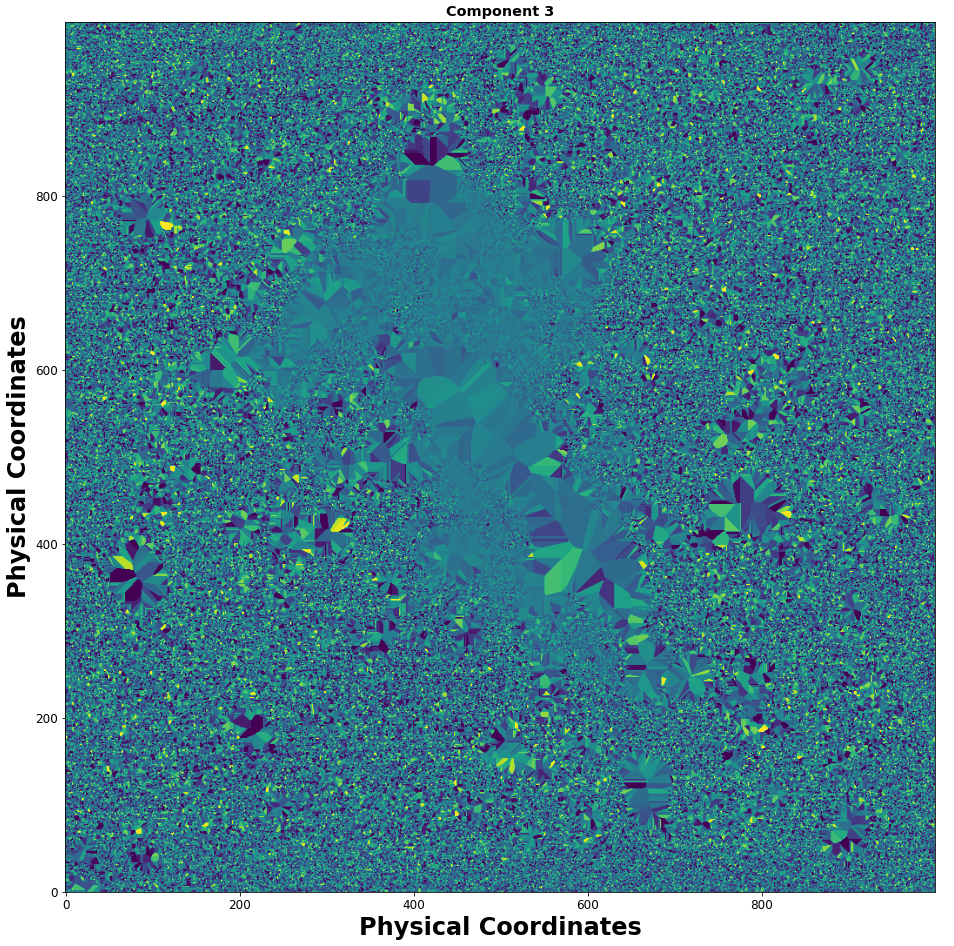}
        \caption{Principal Component 3}
        \hfill
    \end{subfigure}

    \caption{Interpolation of principal component coefficients using nearest neighbor interpolation. The coefficients have been scaled to unity; yellow implies a coefficient value near 1, and dark blue implies a value near 0. Although the background and source pixels are shown in the same color, they are distinguishable by the smoothness of the coefficients in the source regions. For example, regions of background pixels are remarkable in the presence of noise in the coefficient values.
    We note that these coefficients are unitless.}
    \label{fig:nearest-interpolation}
\end{figure*}

These plots (figure \ref{fig:nearest-interpolation}) show the results of using linear interpolation as implemented in \texttt{scipy.griddata(method='nearest')}.

%\section{Perseus Plots}

\begin{comment}
\section{Nonlinear Reconstruction of Spectra}\label{app:nonlinear}
%Using PCA, we assume an adequate linear decomposition of the input spectrum; however, this linearity is not guaranteed. 
When using PCA, we explicity assume that the background spectra are linear combinations of different components in addition to the mean; however, we are not guaranteed that this assumption of linearity is true.
To explore this, we construct an autoencoder that deconstructs the input spectrume into a non-linear latent space (\citealt{hinton_reducing_2006}, \citealt{melchior_autoencoding_2022}; \citealt{ma_spectral-spatial_2016}; \citealt{yang_autoencoder_2015}). We adopt a simple convolutional neural network with two convolutional layers followed by a single fully connected layer that acts as the latent space vector; the decoding is therefore done using two more convolutional layers. We tested several different hyper-parameter configurations (i.e., the size and number of the convolutional kernels, the size of the latent space, and the more traditional hyperparameters such as the learning rate). Our findings indicate that the autoencoder does not achieve a better reconstruction error (i.e., the mean-squared error between the original spectrum and its reduced-dimensionality reconstruction) than the linear principal component decomposition. 
Therefore, the assumption of linearity holds and a PCA decomposition is adequate.
\end{comment}
%%%%%%%%%%%%%%%%%%%%%%%%%%%%%%%%%%%%%%%%%%%%%%%%%%

% Don't change these lines
\bsp	% typesetting comment
\label{lastpage}
\end{document}